\documentclass[superscriptaddress, twocolumn, reprint, citeautoscript, prb]{revtex4-2}
\usepackage{amsmath, amssymb, graphicx, bm}
\usepackage{xr-hyper}
\usepackage[colorlinks=true,citecolor=blue,urlcolor=blue]{hyperref}
\usepackage{xcolor, soul}
\usepackage{gensymb}
\usepackage{float}
\usepackage{bbm}
\usepackage{lipsum}
\usepackage{amsfonts,graphicx}
\usepackage{color}
\usepackage{natbib}
\usepackage{bm}
\usepackage{comment}
\usepackage{enumerate}
\usepackage{booktabs}
\usepackage{multirow}
\usepackage[version=4,arrows=pgf-filled,
textfontname=sffamily,
mathfontname=mathsf]{mhchem}
\usepackage{tabularx}

\begin{document}
\setcitestyle{super}

\title{\textit{De novo} emergence of metabolically active protocells}

\author{Nayan Chakraborty}
\email{nayanc@ncbs.res.in}
\affiliation{National Centre for Biological Sciences (TIFR), Bangalore, India}
\author{Shashi Thutupalli}
\email{shashi@ncbs.res.in}
\affiliation{National Centre for Biological Sciences (TIFR), Bangalore, India}
\affiliation{International Centre for Theoretical Sciences (TIFR), Bangalore, India}

\date{\today}

\begin{abstract}

A continuous route from a disordered soup of simple chemical feedstocks to a functional protocell -- a compartment that metabolizes, grows, and propagates -- remains elusive~\cite{oparin1953origin, Haldane1929, ganti2003principles, rasmussen2003bridging, ruiz-mirazo_2014, jenewein2025concomitant}. Here, we show that a homogeneous aqueous chemical mixture containing phosphorus, iron, molybdenum salts and formaldehyde spontaneously self-organizes into compartments that couple robust non-equilibrium chemical dynamics to their own growth. These structures mature to a sustained, dissipative steady state and support an organic synthetic engine, producing diverse molecular species including many core biomolecular classes. Internal spherules that are themselves growth-competent are produced within the protocells, establishing a rudimentary mode of self-perpetuation. The chemical dynamics we observe in controlled laboratory conditions also occur in reaction mixtures exposed to natural day-night cycles. Strikingly, the morphology and chemical composition of the protocells in our experiments closely resemble molybdenum-rich microspheres recently discovered in current oceanic environments~\cite{shoham_2024}. Our work establishes a robust, testable route to \textit{de novo} protocell formation. The emergence of life-like spatiotemporal organization and chemical dynamics from minimal initial conditions is more facile than previously thought and could be a recurring natural phenomenon~\cite{lehmann_2007, martin_2008, hazen_2010}.

\end{abstract}

\maketitle

The transition from abiotic chemistry to the living state requires the formation of spatio-temporally organized structure and dynamics within high-entropy chemical mixtures. Many studies have elucidated possible routes for the synthesis of life's chemical building blocks from simple feedstocks~\cite{butlerow1861bildung, miller1953production, bahadur1954photosynthesis, bahadur1958photosynthesis, oro1960synthesis, sutherland_2009, patel_2015}. These have involved chemistries and conditions which are often incompatible with each other, precluding a continuous route from these components to a functional, self-organizing compartment. On the other hand, starting mixtures of sufficiently complex and information-rich biomolecules -- what may be termed `synthetic biology approaches' -- have been shown to organize into functional protocell-like dynamical states~\cite{fox1958thermal, fox1959production, luisi1989self, szostak_2001, hanczyc2003experimental, Noireaux2004, mansy_2008, schwille2018maxsynbio, morrow2019chemically, cho_2024, katla2025self, wenisch2025toward}.

Both these frameworks, however, bypass the central question of continuity: how can a functional protocell emerge spontaneously from a simple feedstock mixture? Resolving the discontinuity requires a chemical system that harnesses an energy source to drive non-equilibrium chemical complexification while simultaneously generating its own physical boundary~\cite{prigogine1985self, ganti2003chemoton}. Bahadur and Ranganayaki claimed that sunlight, coupled with the catalytic versatility and redox potential of the transition metal molybdenum and iron~\cite{bahadur1958photosynthesis, benner2018mineral}, offers a potent thermodynamic driver for such self-organization \textit{i.e.} the photo-catalytic transformation of simple chemical mixtures into protocell-like structures capable of growth and chemical synthesis~\cite{ bahadur1964synthesis1, bahadur1964synthesis2, bahadur1964synthesis3}. Inspired by this, we developed a minimal, well-defined system that enables quantitative, mechanistic tests and rigorous characterization of such coupling. 

\begin{figure*}[t!]
\centering
\includegraphics[width=0.85\textwidth]{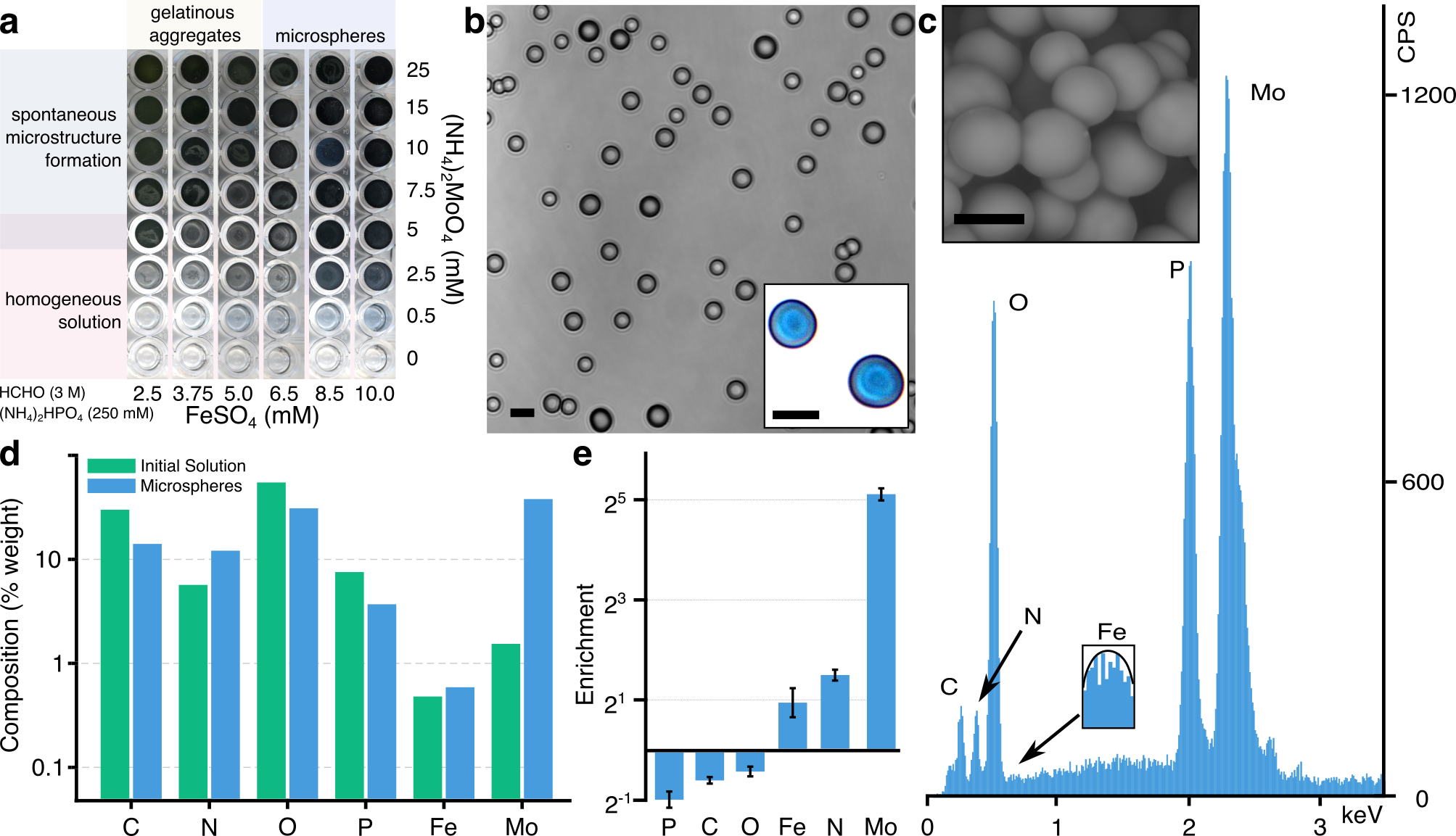}
\caption{\textbf{Self-organization of molybdenum-rich microspheres requires four simple chemical precursors and results from selective elemental sequestration.} \textbf{a}. Formation of microspheres is contingent on the presence of all four primary feedstocks: \ce{(NH4)2MoO4}, \ce{FeSO4}, \ce{HCHO}, \ce{(NH4)2HPO4}, suitably acidified (pH $\sim$ 2; details in the Supplementary Information). Sphere formation is only observed in the complete mixture in a region marked by the blue box. \textbf{b}. Optical micrograph of uniform microspheres. Inset: True color image highlighting the characteristic blue of the spheres. Scale bars, 2 $\mu$m. \textbf{c}. Energy-dispersive X-ray (EDX) spectroscopy showing the elemental composition of the spheres. Inset: Scanning electron microscopy (SEM) image showing the spherical particles on which the spectroscopy was performed. Scale bar, 2 $\mu$m. \textbf{d, e}. The spheres form via a selective elemental incorporation quantified by an enrichment factor (defined as the $\log_2$ fold change in elemental weight fractions; mean $\pm$ SEM, n = 5 replicates). This reveals an enrichment of molybdenum (\ce{Mo}), nitrogen (\ce{N}) and iron (\ce{Fe}) alongside a relative depletion of carbon (\ce{C}), oxygen (\ce{O}) and phosphorus (\ce{P}). }
\label{fig:fig1}
\end{figure*}

We start by incorporating key biogenic elements (\ce{CHNOPS}) and transition metal catalysts (\ce{Fe}, \ce{Mo}) capable of colloidal self-assembly~\cite{liu2003self, miras2020spontaneous} to achieve biosynthesis coupled with structure formation and dynamics. We chose the specific molecular sources of carbon, nitrogen, sulfur and phosphorus based on their preponderance in prebiotic environments~\cite{cleaves2008prebiotic, todd2022sources} and transition metals given their capacity for self-assembly and facile redox coupling~\cite{bahadur1958photosynthesis, benner2018mineral}. Our minimal aqueous mixture was composed of four simple compounds: formaldehyde (\ce{HCHO}), diammonium molybdate (\ce{(NH4)2MoO4}), ferrous sulfate (\ce{FeSO4}), and diammonium hydrogen phosphate (\ce{(NH4)2HPO4}). By systematically varying the starting concentrations and the pH of this feedstock mixture (Fig.~\ref{fig:fig1}\textbf{a}; Methods and Supplementary Information), we examined the resulting self-organization. We discovered that a wide range of conditions resulted in the spontaneous formation of blue micron-sized spherical particles within 1-2 hours of starting the reaction (Fig.~\ref{fig:fig1}\textbf{b}). The formation of the microspheres is strictly contingent on the presence of all four components; removing any of them resulted in a clear, featureless solution.

To understand the composition of the microspheres, we performed energy-dispersive X-ray (EDX) spectroscopy on particles formed approximately 4 hours after initiating the chemical reaction (Fig.~\ref{fig:fig1}\textbf{c}, Methods and Supplementary Information). The analysis revealed that while the particles are comprised of the feedstock elements (\ce{C, N, O, P, Fe, Mo}), their stoichiometric ratios are distinct from those in the bulk solution (Fig.~\ref{fig:fig1}\textbf{d}). We quantified this deviation by computing an enrichment factor (Methods), which confirmed a non-stoichiometric pattern of incorporation (Fig.~\ref{fig:fig1}\textbf{e} and Methods). Molybdenum and nitrogen were strongly enriched, while carbon, oxygen, and phosphorus were relatively depleted in the microspheres. This selective sequestration indicates that the particles form via a chemically biased self-organization process and not through simple precipitation. The strong enrichment of molybdenum is consistent with the characteristic blue hue of the particles driven by a redox process~\cite{muller2000soluble} (Fig.~\ref{fig:fig1}\textbf{b}, inset).

\begin{figure}[t!]
\centering
\includegraphics[width=\columnwidth]{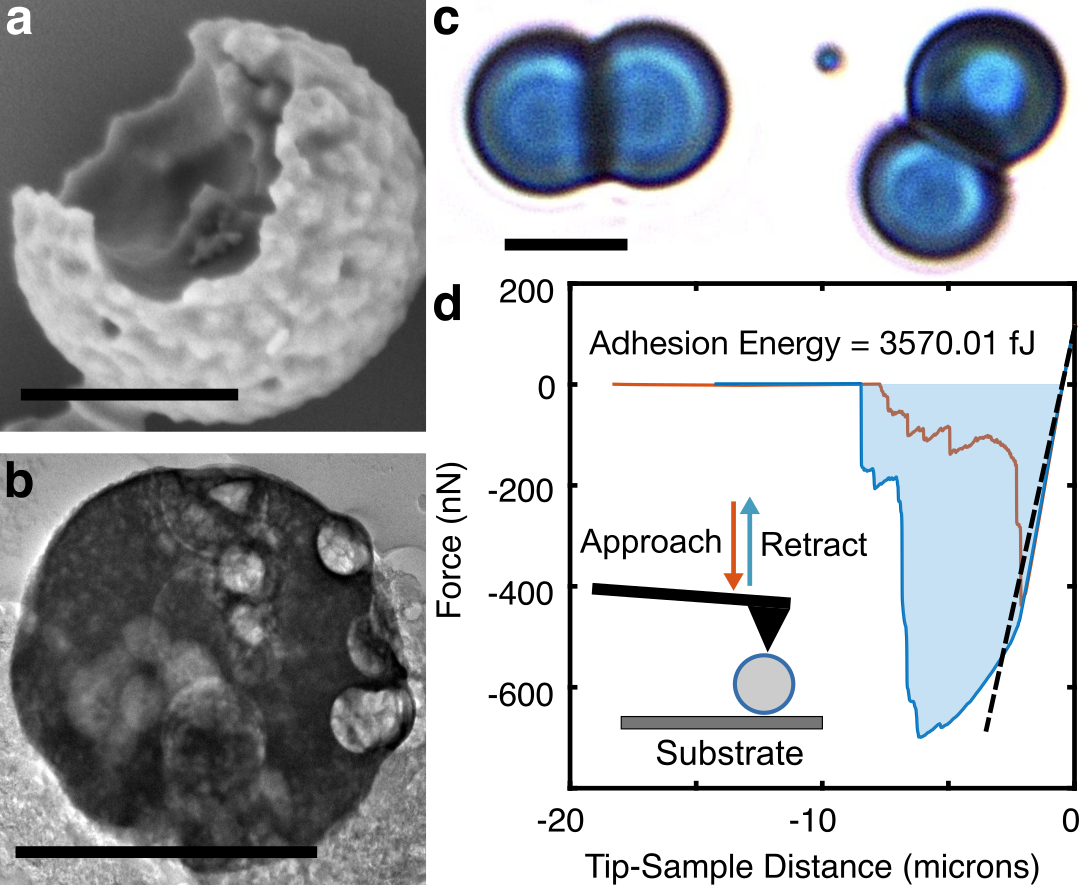}
\caption{\textbf{The microspheres are soft, sticky compartments with heterogeneous interiors.} \textbf{a}. SEM reveals an outer shell, indicating that the microspheres are characterized by a boundary wall and a lumen. Scale bar: $2~\mu m$. \textbf{b}. TEM shows contrast due to electron density differences, pointing to internal heterogeneity and sphere-like inclusions. Scale bar: $2~\mu m$. \textbf{c}. True color images of representative microsphere pairs exhibiting a stable contact indicating deformability and integrity of the boundary against fusion. Scale bar: $2~\mu m$. \textbf{d}. AFM measurements confirm mechanical softness and stickiness of the microspheres. The dashed line shows a guide to the eye, representing the stiffness of the AFM cantilever ($\sim$ 0.23 N/m).}
\label{fig:fig2}
\end{figure}

High-resolution imaging confirms that these particles are compartments with distinct structural features. Scanning electron microscopy (SEM) reveals a well-defined boundary wall enclosing a hollow lumen (Fig.~\ref{fig:fig2}\textbf{a}, Methods). Transmission electron microscopy (TEM) further shows a spatially heterogeneous interior containing smaller, spherical inclusions (Fig.~\ref{fig:fig2}\textbf{b}, Methods), indicated by the contrast due to variations in the electron density. Laser-induced perforation of the boundary wall resulted in the expulsion of the lumenal contents, indicating their liquid-like nature (Supplementary Information and Supplementary Video 1). These compartments are soft, adhesive shells: contacting compartments deform at the interface, forming a stable contact area while resisting fusion (Fig.~\ref{fig:fig2}\textbf{c}, Methods). Atomic force microscopy (AFM) measurements provide quantitative confirmation of these properties, revealing a low stiffness and high adhesion energy (Fig.~\ref{fig:fig2}\textbf{d}, Methods); the particles exhibit viscoelasticity, with signs of aging of the shells as they harden over time (Supplementary Information).

\begin{figure*}[t!]
\centering
\includegraphics[width=0.85\textwidth]{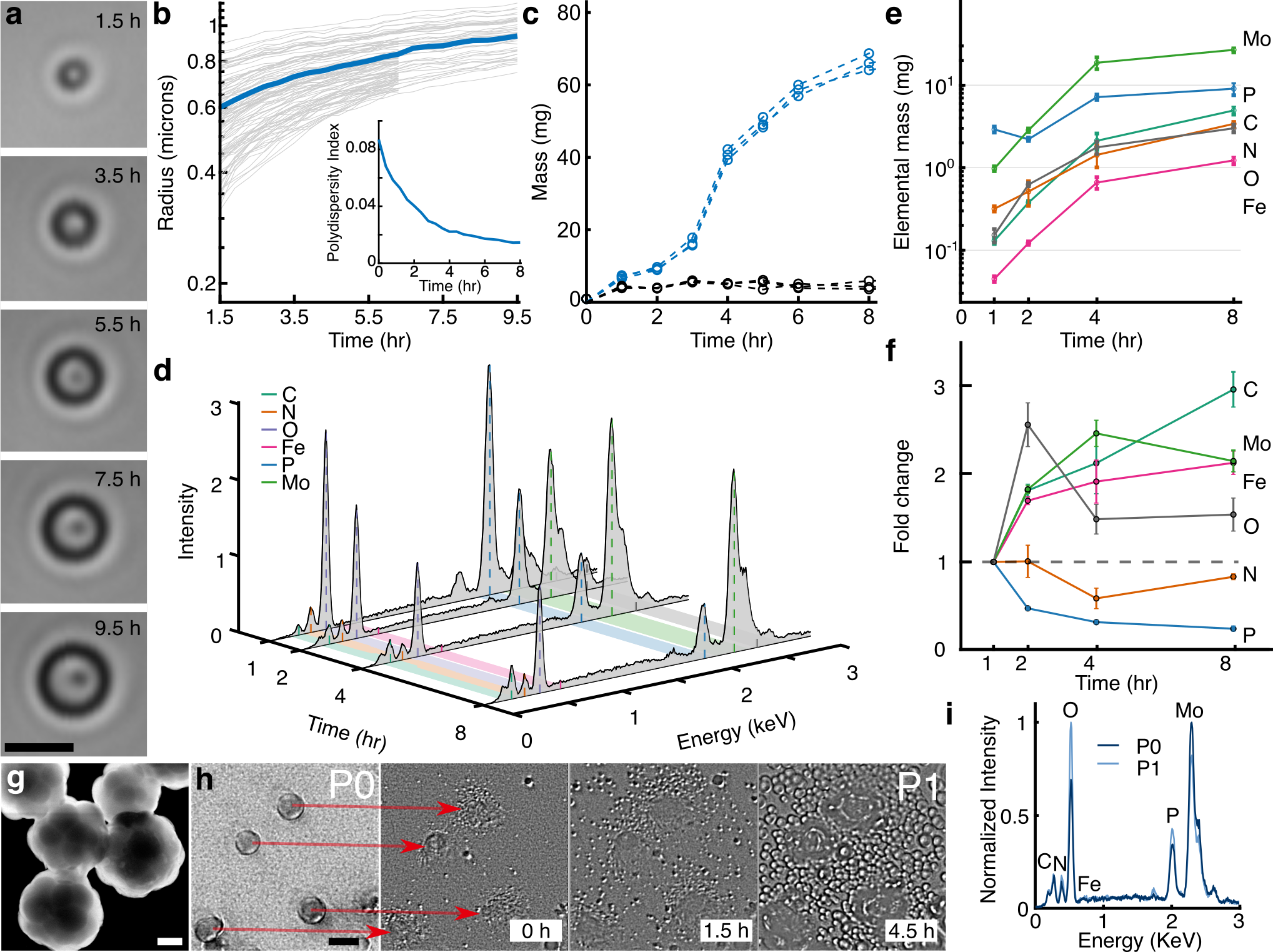}
\caption{\textbf{The compartments grow and mature, and have a capacity for self-perpetuation.} \textbf{a}. Time-lapse brightfield microscopy snapshots showing the growth of a representative compartment. Scale bar, 2~$\mu m$. \textbf{b}. The mean radius (blue line) of the compartment population increases over time (individual compartment traces are shown in light grey). Inset: The polydispersity index (PDI) of the population decreases with time. \textbf{c}. The total mass of the compartment population increases over time, confirming material transfer from the solution. \textbf{d}. Time-resolved EDX spectra showing the dynamic elemental composition of the compartments. \textbf{e}. The absolute mass of all constituent elements increases throughout the growth process. \textbf{f}. Fold change in elemental weight fractions reveals a dynamic compositional evolution; the composition stabilizes after $\sim$4 hours. \textbf{g}. Transmission electron microscopy (TEM) images of compartments, 48 hours post formation, showing internal spherical structures. Scale bar, 2~$\mu m$. \textbf{h}. Time-lapse microscopy showing the growth of seed particles released from the lumen of mature compartments (indicated by red arrows), 12 hours after formation, when exposed to the feedstock solution. Scale bar, 5~$\mu m$. \textbf{i}. EDX spectrum of the seed microspheres (P1) resulting from growth, demonstrating compositional similarity to the mature compartments (P0).}
\label{fig:fig3}
\end{figure*}

The compartments are the result of a dynamic process of nucleation, growth and maturation into a dynamic non-equilibrium chemical state. Starting from the nucleation of nanoscopic structures, typically 1-2 hours from initiation, the compartments consistently increase in radius over time (Fig.~\ref{fig:fig3}\textbf{a, b} and Supplementary Video 2), corresponding to a net transfer of mass from the bulk solution to the particle population (Fig.~\ref{fig:fig3}\textbf{c}, Methods). The growth is size-dependent, as evidenced by a slowing down of the growth rate as a function of the compartment radius, a kinetic signature of a surface-catalyzed mechanism, distinct from simple precipitation. This growth mechanism results in an overall decrease in the size polydispersity of the compartment population (Fig.~\ref{fig:fig3}\textbf{b}, inset and Supplementary Information). The physical growth of the compartments is coupled to their chemical maturation. Time-resolved EDX analysis (Fig.~\ref{fig:fig3}\textbf{d}) shows that while the absolute mass of all constituent elements increases (Fig.~\ref{fig:fig3}\textbf{e}), their relative proportions within the compartment evolve dynamically before reaching a steady composition after $\sim$4 hours (Fig.~\ref{fig:fig3}\textbf{f}).

Strikingly, the internal chemical activity is coupled to the production of structures that are themselves capable of growth. A few hours after formation, we found that smaller, nanoscopic inclusions develop within the lumen of mature compartments (also seen in Supplementary Video 1). Since these are difficult to track via optical microscopy, we performed high intensity TEM imaging on the compartments 48 hours after their formation to  reveal larger spherical structures within (Fig.~\ref{fig:fig3}\textbf{g}), presumably resulting from the growth of the aforementioned inclusions. To test this idea, we ruptured the wall of compartments 12 hours after their formation to release the small inclusions into the external fluid medium (Methods) to find that they grow over time into larger spherical particles (marked as P1 in Fig.~\ref{fig:fig3}\textbf{h} and Supplementary Video 3). Further, we analysed the elemental composition of the newly formed structures (P1) and found them to be similar to the compartments (P0) from which they were originally released (Fig.~\ref{fig:fig3}\textbf{i}). Taken together, these observations raise the possibility of a rudimentary mode of self-perpetuation: the internal processes generate growth-competent structures that can serve as seeds for subsequent generations. 

We next asked whether the compartment formation is accompanied by sustained, non-equilibrium chemical activity and organic synthesis. To quantify the energetic signature of the reaction, we performed isothermal calorimetry on sealed samples maintained at $30\,^{\circ}\mathrm{C}$ (Methods and Supplementary Information). The complete reaction mixture exhibited sustained exothermic activity over extended durations (up to 21~days in our measurements), whereas control mixtures lacking formaldehyde produced substantially less heat (Fig.~\ref{fig:fig4}\textbf{a} and more details in Supplementary Information). This prolonged energy release was accompanied by systematic changes in bulk solution acidity over time (Methods and Supplementary Information), consistent with continued chemical turnover in the reaction mixture rather than a transient burst of activity.

Notably, the calorimetry measurements were performed in the dark. We therefore repeated the compartment formation and growth assays under dark conditions and confirmed that the compartments form and grow without any illumination, with kinetics comparable to those observed under ambient laboratory lighting (Supplementary Information). However, to impose a controlled and reproducible external energy input for long-term chemical characterization, we performed subsequent molecular analyses under a solar-spectrum simulator (AM1.5G; 1~Sun-equivalent irradiance), with reactions maintained at $25\,^{\circ}\mathrm{C}$ in a class~10000 clean-room environment (Fig.~\ref{fig:fig4}\textbf{b}, Methods). These conditions enable direct comparison across experiments and facilitate longer-term incubation without confounding fluctuations in illumination or temperature.

To trace carbon flux from the C$_1$ feedstock into newly formed products, we supplied $^{13}$C-labeled formaldehyde and monitored the evolving chemical mixture using $^{13}$C NMR spectroscopy (Methods and Supplementary Information). Using 99\% $^{13}$C-labeled formaldehyde as the sole carbon source, time-resolved spectra revealed the progressive emergence and growth of multiple new $^{13}$C chemical-shift peaks, concomitant with depletion of the formaldehyde signal (Fig.~\ref{fig:fig4}\textbf{c}; Supplementary Information). Quantitative analysis of the NMR peak integrals showed a systematic reduction in the total formaldehyde pool together with an increase in newly formed $^{13}$C-bearing products, indicating sustained C$_1$-source organic synthesis.

\begin{figure*}
\centering
\includegraphics[width=0.85\textwidth]{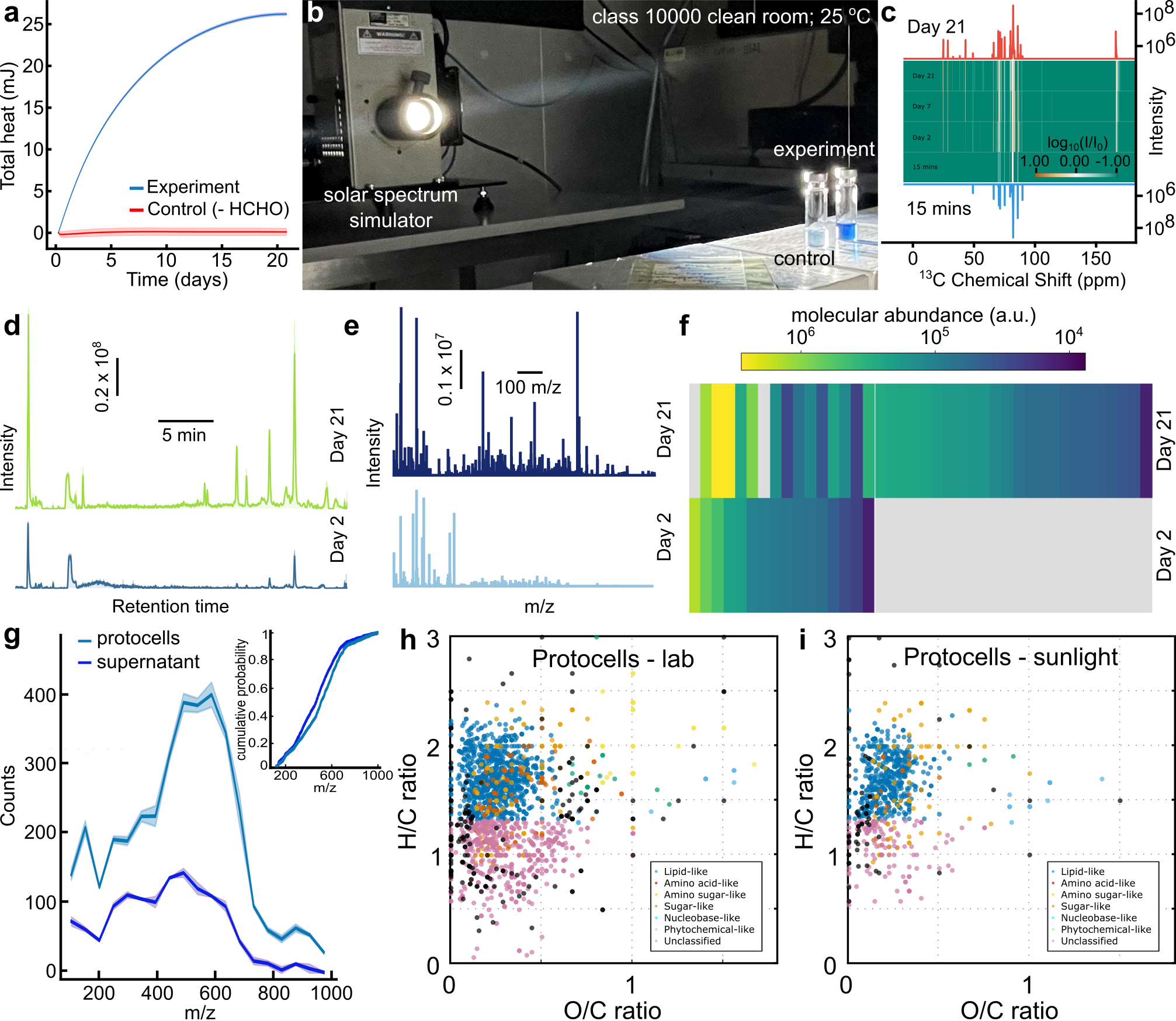}
\caption{\textbf{The compartments are metabolically active chemical reactors that synthesize diverse biomolecular precursors.} 
\textbf{a}. Isothermal calorimetry shows sustained exothermic activity in the complete reaction mixture (blue) compared to control mixtures lacking formaldehyde (red) over 21 days. 
\textbf{b}. Laboratory setup of the chemical reaction mixture exposed to a steady solar-spectrum simulating light source (AM1.5G; 1 Sun-equivalent irradiance). The reaction mixtures were infused with \textsuperscript{13}C-labeled formaldehyde.
\textbf{c}. Quantitative analysis of 1-D \textsuperscript{13}C Nuclear Magnetic Resonance (NMR) spectra for the complete reaction mixture using 99\% \textsuperscript{13}C-labeled formaldehyde shows the time-dependent emergence and increasing intensity of new molecular peaks.
\textbf{d}. Liquid Chromatography (LC) analysis of the complete reaction mixture shows the appearance of numerous new peaks from day 2 to day 21.
\textbf{e}. Mass Spectrometry (MS) plots of m/z versus intensity for the complete reaction mixture reveal the progressive appearance of many new peaks with increasing mass values from day 2 to day 21.
\textbf{f}. Mass Spectrometry (MS) analysis of the complete reaction mixture shows the time-dependent emergence of new molecular species, along with a turnover in their molecular abundance from day 2 to day 21.
\textbf{g}. Comparison of the molecular-weight distributions in the compartment-enriched phase and the surrounding supernatant phase, showing that the compartments are enriched in higher-molecular-weight products. Experiments for this analysis were carried out by spiking with 1\% \textsuperscript{13}C-labeled formaldehyde. Inset shows the plots for cumulative distribution function (CDF) for the two phases.
Van Krevelen diagrams (H/C versus O/C) along with Multidimensional Stoichiometric Compound Classification (MSCC)~\cite{krevelen1950graphical, kim2003graphical, rivas2018moving} for all detected organic products shown for the protocell phase exposed to \textbf{h}. lab solar simulator and \textbf{i}. natural day-night cycles.}
\label{fig:fig4}
\end{figure*}

We next assessed the diversification of products using liquid chromatography and high-resolution mass spectrometry (LC--MS; Methods). LC analysis of the complete mixtures revealed the appearance of numerous additional peaks at day 21 compared to day 2, spanning a broad range of retention times and therefore consistent with products of diverse polarities (Fig.~\ref{fig:fig4}\textbf{d}). Concomitantly, high-resolution MS revealed a progressive increase in spectral complexity, including the appearance of many new features extending to higher $m/z$ values over time (Fig.~\ref{fig:fig4}\textbf{e}). Beyond this net increase in complexity, tracking the time-dependent abundances of individual $m/z$ features demonstrated non-trivial chemical dynamics: while some species accumulate, others are depleted as new features appear, highlighting ongoing turnover and transformation rather than monotonic build-up (Fig.~\ref{fig:fig4}\textbf{f}, Methods and Supplementary Information). 

To determine whether reaction products are spatially localized within the compartments, we used larger volumes of reaction mixtures spiked with 1 \% $^{13}$C fraction. We analysed the compartment-enriched fraction separately from the surrounding supernatant (Methods and Supplementary Information). The compartments displayed a pronounced enrichment of higher-$m/z$ species relative to the supernatant, as quantified by their molecular distributions and corresponding cumulative distribution functions (Fig.~\ref{fig:fig4}\textbf{g}). These results indicate that the synthesis of higher-$m/z$ products occurs within, or proximal to, the compartments. Finally, to obtain a coarse-grained view of the chemical diversity sampled by compartment-associated products, we applied elemental-ratio--based classification~\cite{krevelen1950graphical, kim2003graphical, rivas2018moving} (Fig.~\ref{fig:fig4}\textbf{h}). A subset of detected molecular species falls within stoichiometric regions consistent with lipid-like, amino-acid/peptide-like, and carbohydrate-like classes (candidate molecular identification across these classes is shown in the Supplementary Information). Importantly, these global chemical signatures were robust across environments: long-term reaction mixtures exposed directly to natural outdoor day--night cycles for two months yielded molecular distributions similar to those obtained under laboratory solar simulation (Fig.~\ref{fig:fig4}i).

\begin{figure*}[t!]
\centering
\includegraphics[width=1.6\columnwidth]{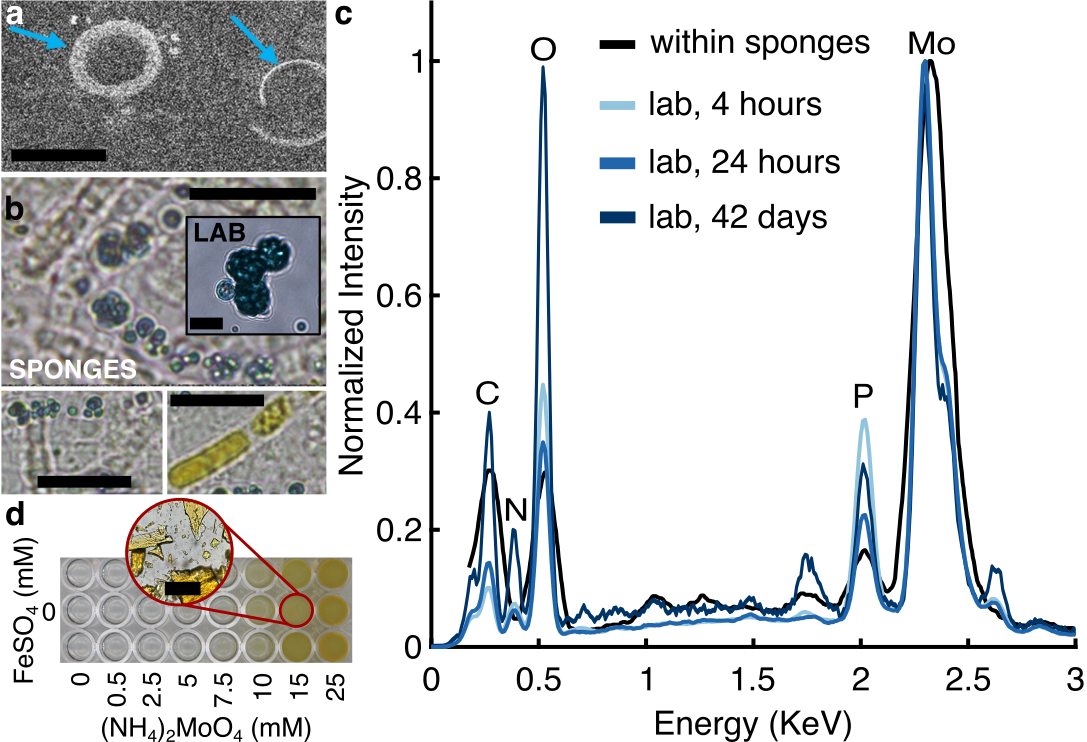}
\caption{\textbf{The synthetic compartments exhibit similar morphology and chemical composition to naturally occurring molybdenum-rich microspheres in the ocean.} \textbf{a}. SEM and \textbf{b}. brightfield microscopy images of the natural microspheres (referred to as ``blue vacuoles'') found within the Indo-Pacific sponge \textit{Theonella conica} showing their hollow, shell-like morphology~\cite{shoham_2024}. All scale bars, 2~$\mu m$. Inset in panel \textbf{b} shows protocells obtained from our experiments 42 days after initiation of the reaction in laboratory conditions described in Fig.~\ref{fig:fig4}. Scale bar, 1~$\mu m$. \textbf{c}. The EDX spectra of laboratory-formed compartments (4 hours, light blue; 24 hours, dark blue and 42 days, darker blue) are overlaid with the spectra obtained from oceanic microspheres (black)~\cite{shoham_2024}. \textbf{d}. Image from the chemical phase space of our experiment showing the formation of yellow structures, a color polymorphism also observed in the natural system (as seen in panel \textbf{b}, lower right image) and attributed to the redox chemistry of molybdenum~\cite{saji2012molybdenum}. Scale bar, 1~$\mu m$. Images in panels \textbf{a} and \textbf{b} and data in panel \textbf{c} are reproduced with permission from the \textit{American Association for the Advancement of Science}.}
\label{fig:fig5}
\end{figure*}

Taken together, our results demonstrate a continuous route from simple feedstocks to protocells that exhibit proto-metabolic behaviour: compartments that actively shape their own composition while sustaining organic synthesis. This is more than chemistry in bulk solution or passive precipitation: the system remains in a long‑lived, energy‑dissipating non‑equilibrium regime with continued feedstock consumption and product turnover, rather than relaxing after a transient burst. Crucially, this turnover is spatially organized by the compartments, which selectively sequester and enrich complex reaction products; this localized chemistry is coupled to compartment growth, compositional maturation, and the generation of growth-competent internal seeds. Such coupling is the hallmark of proto-metabolism~\cite{morowitz2007energy, muchowska2020nonenzymatic}, distinct from the enzyme-regulated networks of extant life. Although our feedstock selection was motivated by prebiotic plausibility, such molecular species remain available in modern geological settings such as mineral-rich ocean beds and hydrothermal vents~\cite{helz2004molybdenum, martin_2008, inaba2018primary, evans2023molybdenum, takahagi2023extreme, randolph2023phosphate}, where ``environmental scaffolding''~\cite{classics1966chemiosmotic, kauffman1971cellular, corliss1990hot, lane2015vital, harrison2023life} could sustain the non-equilibrium dynamics observed here.

Recently identified micron-sized molybdenum-rich microspheres described as hollow ``blue vacuoles'', within the Indo-Pacific sponge \textit{Theonella conica}~\cite{shoham_2024} show a remarkable convergence with the protocells in our experiments (Fig.~\ref{fig:fig5}\textbf{a, b}). The morphology of the oceanic structures is consistent with the shell-like structure of our synthetic protocells (Fig.~\ref{fig:fig2}\textbf{a} and Supplementary Information). Further, the elemental composition of these naturally occurring structures reveals a remarkable similarity: both the natural and synthetic compartments are dominated by a molybdenum peak and show significant amounts of carbon, oxygen, and phosphorus (Fig.~\ref{fig:fig5}\textbf{c}). Notably, the time evolving composition of the synthetic protocells converges closely to the elemental composition of the naturally occurring structures (Fig.~\ref{fig:fig5}\textbf{c}). The convergence extends further: in addition to the ``blue vacuoles'', Shoham \textit{et al.} report yellow microstructures, attributed to different molybdenum--associated chemical compositions. Indeed, we observe similar yellow structures in our experiments, a color polymorphism occurring in a chemical space proximal to that where the spherical compartments form (Fig.~\ref{fig:fig5}\textbf{d} and Supplementary Information).

More broadly, our findings suggest a resolution to the discontinuity problem: under suitable conditions, dissipation-driven chemical complexification and boundary formation can co-emerge and stabilize one another as a single non-equilibrium process~\cite{prigogine1985self, ganti2003chemoton}. Posed this way, protocell formation becomes a feedback problem in which catalytic chemistry generates a boundary that selectively partitions matter, while that boundary in turn biases the chemistry towards growth and molecular diversification. Because the minimal system we introduce is robust and experimentally tractable, it provides a quantitative platform to identify the underlying chemical engine and to test whether the evolving reaction network can approach autocatalytic closure and inheritance--like propagation~\cite{kauffman1995home, bissette2013mechanisms}. Mapping how such metal-rich compartments recruit additional functions such as selective permeability, amphiphilic membranes, and templating polymers will help connect simple growth--competent protocells to extant cellular organization. Further, this framing of non-equilibrium organometallic microcompartments that localize dissipation and concentrate evolving organics points to generic physicochemical signatures of ``life-adjacent'' organization ~\cite{davies1999fifth, kauffman2000investigations}. This could therefore help broaden search for life on Earth and in extraterrestrial environments where similar chemistries operate~\cite{hurowitz2025redox, schwieterman2018exoplanet, national2019astrobiology, jia2018evidence, liao2020heat}. In particular, if metabolically active compartmentalization can appear robustly across conditions, then the operational criteria used to prioritize environments for biosignature discovery may be broader than typically assumed~\cite{shock2007quantitative, cockell2024sustained}.

\newpage

\bibliography{04_references}

\begin{thebibliography}{68}%
\makeatletter
\providecommand \@ifxundefined [1]{%
 \@ifx{#1\undefined}
}%
\providecommand \@ifnum [1]{%
 \ifnum #1\expandafter \@firstoftwo
 \else \expandafter \@secondoftwo
 \fi
}%
\providecommand \@ifx [1]{%
 \ifx #1\expandafter \@firstoftwo
 \else \expandafter \@secondoftwo
 \fi
}%
\providecommand \natexlab [1]{#1}%
\providecommand \enquote  [1]{``#1''}%
\providecommand \bibnamefont  [1]{#1}%
\providecommand \bibfnamefont [1]{#1}%
\providecommand \citenamefont [1]{#1}%
\providecommand \href@noop [0]{\@secondoftwo}%
\providecommand \href [0]{\begingroup \@sanitize@url \@href}%
\providecommand \@href[1]{\@@startlink{#1}\@@href}%
\providecommand \@@href[1]{\endgroup#1\@@endlink}%
\providecommand \@sanitize@url [0]{\catcode `\\12\catcode `\$12\catcode `\&12\catcode `\#12\catcode `\^12\catcode `\_12\catcode `\%12\relax}%
\providecommand \@@startlink[1]{}%
\providecommand \@@endlink[0]{}%
\providecommand \url  [0]{\begingroup\@sanitize@url \@url }%
\providecommand \@url [1]{\endgroup\@href {#1}{\urlprefix }}%
\providecommand \urlprefix  [0]{URL }%
\providecommand \Eprint [0]{\href }%
\providecommand \doibase [0]{https://doi.org/}%
\providecommand \selectlanguage [0]{\@gobble}%
\providecommand \bibinfo  [0]{\@secondoftwo}%
\providecommand \bibfield  [0]{\@secondoftwo}%
\providecommand \translation [1]{[#1]}%
\providecommand \BibitemOpen [0]{}%
\providecommand \bibitemStop [0]{}%
\providecommand \bibitemNoStop [0]{.\EOS\space}%
\providecommand \EOS [0]{\spacefactor3000\relax}%
\providecommand \BibitemShut  [1]{\csname bibitem#1\endcsname}%
\let\auto@bib@innerbib\@empty
\bibitem [{\citenamefont {Oparin}(1953)}]{oparin1953origin}%
  \BibitemOpen
  \bibfield  {author} {\bibinfo {author} {\bibfnamefont {A.~I.}\ \bibnamefont {Oparin}},\ }\bibfield  {title} {\bibinfo {title} {The origin of life},\ }\href@noop {} {\  (\bibinfo {year} {1953})}\BibitemShut {NoStop}%
\bibitem [{\citenamefont {Haldane}(1929)}]{Haldane1929}%
  \BibitemOpen
  \bibfield  {author} {\bibinfo {author} {\bibfnamefont {J.}~\bibnamefont {Haldane}},\ }\bibfield  {title} {\bibinfo {title} {{The origin of life}},\ }\href@noop {} {\bibfield  {journal} {\bibinfo  {journal} {Ration. Annu.}\ }\textbf {\bibinfo {volume} {148}},\ \bibinfo {pages} {3} (\bibinfo {year} {1929})}\BibitemShut {NoStop}%
\bibitem [{\citenamefont {G{\'a}nti}(2003{\natexlab{a}})}]{ganti2003principles}%
  \BibitemOpen
  \bibfield  {author} {\bibinfo {author} {\bibfnamefont {T.}~\bibnamefont {G{\'a}nti}},\ }\href@noop {} {\emph {\bibinfo {title} {The principles of life}}}\ (\bibinfo  {publisher} {Oxford University Press},\ \bibinfo {year} {2003})\BibitemShut {NoStop}%
\bibitem [{\citenamefont {Rasmussen}\ \emph {et~al.}(2003)\citenamefont {Rasmussen}, \citenamefont {Chen}, \citenamefont {Nilsson},\ and\ \citenamefont {Abe}}]{rasmussen2003bridging}%
  \BibitemOpen
  \bibfield  {author} {\bibinfo {author} {\bibfnamefont {S.}~\bibnamefont {Rasmussen}}, \bibinfo {author} {\bibfnamefont {L.}~\bibnamefont {Chen}}, \bibinfo {author} {\bibfnamefont {M.}~\bibnamefont {Nilsson}},\ and\ \bibinfo {author} {\bibfnamefont {S.}~\bibnamefont {Abe}},\ }\href@noop {} {\emph {\bibinfo {title} {Protocells: Bridging nonliving and living matter}}}\ (\bibinfo  {publisher} {MIT Press},\ \bibinfo {year} {2003})\BibitemShut {NoStop}%
\bibitem [{\citenamefont {Ruiz-Mirazo}\ \emph {et~al.}(2014)\citenamefont {Ruiz-Mirazo}, \citenamefont {Briones},\ and\ \citenamefont {de~la Escosura}}]{ruiz-mirazo_2014}%
  \BibitemOpen
  \bibfield  {author} {\bibinfo {author} {\bibfnamefont {K.}~\bibnamefont {Ruiz-Mirazo}}, \bibinfo {author} {\bibfnamefont {C.}~\bibnamefont {Briones}},\ and\ \bibinfo {author} {\bibfnamefont {A.}~\bibnamefont {de~la Escosura}},\ }\bibfield  {title} {\bibinfo {title} {Prebiotic systems chemistry: New perspectives for the origins of life},\ }\href@noop {} {\bibfield  {journal} {\bibinfo  {journal} {Chemical Reviews}\ }\textbf {\bibinfo {volume} {114}},\ \bibinfo {pages} {285} (\bibinfo {year} {2014})}\BibitemShut {NoStop}%
\bibitem [{\citenamefont {Jenewein}\ \emph {et~al.}(2025)\citenamefont {Jenewein}, \citenamefont {Ma{\'\i}z-Sicilia}, \citenamefont {Rull}, \citenamefont {Gonz{\'a}lez-Souto},\ and\ \citenamefont {Garc{\'\i}a-Ruiz}}]{jenewein2025concomitant}%
  \BibitemOpen
  \bibfield  {author} {\bibinfo {author} {\bibfnamefont {C.}~\bibnamefont {Jenewein}}, \bibinfo {author} {\bibfnamefont {A.}~\bibnamefont {Ma{\'\i}z-Sicilia}}, \bibinfo {author} {\bibfnamefont {F.}~\bibnamefont {Rull}}, \bibinfo {author} {\bibfnamefont {L.}~\bibnamefont {Gonz{\'a}lez-Souto}},\ and\ \bibinfo {author} {\bibfnamefont {J.~M.}\ \bibnamefont {Garc{\'\i}a-Ruiz}},\ }\bibfield  {title} {\bibinfo {title} {Concomitant formation of protocells and prebiotic compounds under a plausible early earth atmosphere},\ }\href@noop {} {\bibfield  {journal} {\bibinfo  {journal} {Proceedings of the National Academy of Sciences}\ }\textbf {\bibinfo {volume} {122}},\ \bibinfo {pages} {e2413816122} (\bibinfo {year} {2025})}\BibitemShut {NoStop}%
\bibitem [{\citenamefont {Shoham}\ \emph {et~al.}(2024)\citenamefont {Shoham}, \citenamefont {Keren},\ and\ \citenamefont {Ilan}}]{shoham_2024}%
  \BibitemOpen
  \bibfield  {author} {\bibinfo {author} {\bibfnamefont {S.}~\bibnamefont {Shoham}}, \bibinfo {author} {\bibfnamefont {R.}~\bibnamefont {Keren}},\ and\ \bibinfo {author} {\bibfnamefont {M.}~\bibnamefont {Ilan}},\ }\bibfield  {title} {\bibinfo {title} {Out of the blue: Hyperaccumulation of molybdenum in the indo-pacific sponge \textit{Theonella conica}},\ }\href@noop {} {\bibfield  {journal} {\bibinfo  {journal} {Science Advances}\ }\textbf {\bibinfo {volume} {10}},\ \bibinfo {pages} {eadn3923} (\bibinfo {year} {2024})}\BibitemShut {NoStop}%
\bibitem [{\citenamefont {Lehmann}\ \emph {et~al.}(2007)\citenamefont {Lehmann}, \citenamefont {N{\"a}gler}, \citenamefont {Wille}, \citenamefont {Holland},\ and\ \citenamefont {Mao}}]{lehmann_2007}%
  \BibitemOpen
  \bibfield  {author} {\bibinfo {author} {\bibfnamefont {B.}~\bibnamefont {Lehmann}}, \bibinfo {author} {\bibfnamefont {T.~F.}\ \bibnamefont {N{\"a}gler}}, \bibinfo {author} {\bibfnamefont {M.}~\bibnamefont {Wille}}, \bibinfo {author} {\bibfnamefont {H.~D.}\ \bibnamefont {Holland}},\ and\ \bibinfo {author} {\bibfnamefont {J.}~\bibnamefont {Mao}},\ }\bibfield  {title} {\bibinfo {title} {Highly metalliferous carbonaceous shale and early cambrian seawater},\ }\href@noop {} {\bibfield  {journal} {\bibinfo  {journal} {Geology}\ }\textbf {\bibinfo {volume} {35}},\ \bibinfo {pages} {403} (\bibinfo {year} {2007})}\BibitemShut {NoStop}%
\bibitem [{\citenamefont {Martin}\ \emph {et~al.}(2008)\citenamefont {Martin}, \citenamefont {Baross}, \citenamefont {Kelley},\ and\ \citenamefont {Russell}}]{martin_2008}%
  \BibitemOpen
  \bibfield  {author} {\bibinfo {author} {\bibfnamefont {W.}~\bibnamefont {Martin}}, \bibinfo {author} {\bibfnamefont {J.}~\bibnamefont {Baross}}, \bibinfo {author} {\bibfnamefont {D.}~\bibnamefont {Kelley}},\ and\ \bibinfo {author} {\bibfnamefont {M.~J.}\ \bibnamefont {Russell}},\ }\bibfield  {title} {\bibinfo {title} {Hydrothermal vents and the origin of life},\ }\href@noop {} {\bibfield  {journal} {\bibinfo  {journal} {Nature Reviews Microbiology}\ }\textbf {\bibinfo {volume} {6}},\ \bibinfo {pages} {805} (\bibinfo {year} {2008})}\BibitemShut {NoStop}%
\bibitem [{\citenamefont {Hazen}\ and\ \citenamefont {Sverjensky}(2010)}]{hazen_2010}%
  \BibitemOpen
  \bibfield  {author} {\bibinfo {author} {\bibfnamefont {R.~M.}\ \bibnamefont {Hazen}}\ and\ \bibinfo {author} {\bibfnamefont {D.~A.}\ \bibnamefont {Sverjensky}},\ }\bibfield  {title} {\bibinfo {title} {Mineral surfaces, geochemical complexities and the origins of life},\ }\href@noop {} {\bibfield  {journal} {\bibinfo  {journal} {Cold Spring Harbor Perspectives in Biology}\ }\textbf {\bibinfo {volume} {2}},\ \bibinfo {pages} {a002162} (\bibinfo {year} {2010})}\BibitemShut {NoStop}%
\bibitem [{\citenamefont {Butlerow}(1861)}]{butlerow1861bildung}%
  \BibitemOpen
  \bibfield  {author} {\bibinfo {author} {\bibfnamefont {A.}~\bibnamefont {Butlerow}},\ }\bibfield  {title} {\bibinfo {title} {Bildung einer zuckerartigen substanz durch synthese},\ }\href@noop {} {\bibfield  {journal} {\bibinfo  {journal} {Justus Liebigs Annalen der Chemie}\ }\textbf {\bibinfo {volume} {120}},\ \bibinfo {pages} {295} (\bibinfo {year} {1861})}\BibitemShut {NoStop}%
\bibitem [{\citenamefont {Miller}(1953)}]{miller1953production}%
  \BibitemOpen
  \bibfield  {author} {\bibinfo {author} {\bibfnamefont {S.~L.}\ \bibnamefont {Miller}},\ }\bibfield  {title} {\bibinfo {title} {A production of amino acids under possible primitive earth conditions},\ }\href@noop {} {\bibfield  {journal} {\bibinfo  {journal} {Science}\ }\textbf {\bibinfo {volume} {117}},\ \bibinfo {pages} {528} (\bibinfo {year} {1953})}\BibitemShut {NoStop}%
\bibitem [{\citenamefont {Bahadur}(1954)}]{bahadur1954photosynthesis}%
  \BibitemOpen
  \bibfield  {author} {\bibinfo {author} {\bibfnamefont {K.}~\bibnamefont {Bahadur}},\ }\bibfield  {title} {\bibinfo {title} {Photosynthesis of amino-acids from paraformaldehyde and potassium nitrate},\ }\href@noop {} {\bibfield  {journal} {\bibinfo  {journal} {Nature}\ }\textbf {\bibinfo {volume} {173}},\ \bibinfo {pages} {1141} (\bibinfo {year} {1954})}\BibitemShut {NoStop}%
\bibitem [{\citenamefont {Bahadur}\ \emph {et~al.}(1958)\citenamefont {Bahadur}, \citenamefont {Ranganayaki},\ and\ \citenamefont {Santamaria}}]{bahadur1958photosynthesis}%
  \BibitemOpen
  \bibfield  {author} {\bibinfo {author} {\bibfnamefont {K.}~\bibnamefont {Bahadur}}, \bibinfo {author} {\bibfnamefont {S.}~\bibnamefont {Ranganayaki}},\ and\ \bibinfo {author} {\bibfnamefont {L.}~\bibnamefont {Santamaria}},\ }\bibfield  {title} {\bibinfo {title} {Photosynthesis of amino-acids from paraformaldehyde involving the fixation of nitrogen in the presence of colloidal molybdenum oxide as catalyst},\ }\href@noop {} {\bibfield  {journal} {\bibinfo  {journal} {Nature}\ }\textbf {\bibinfo {volume} {182}},\ \bibinfo {pages} {1668} (\bibinfo {year} {1958})}\BibitemShut {NoStop}%
\bibitem [{\citenamefont {Or{\'o}}(1960)}]{oro1960synthesis}%
  \BibitemOpen
  \bibfield  {author} {\bibinfo {author} {\bibfnamefont {J.}~\bibnamefont {Or{\'o}}},\ }\bibfield  {title} {\bibinfo {title} {Synthesis of adenine from ammonium cyanide},\ }\href@noop {} {\bibfield  {journal} {\bibinfo  {journal} {Biochemical and Biophysical Research Communications}\ }\textbf {\bibinfo {volume} {2}},\ \bibinfo {pages} {407} (\bibinfo {year} {1960})}\BibitemShut {NoStop}%
\bibitem [{\citenamefont {Powner}\ \emph {et~al.}(2009)\citenamefont {Powner}, \citenamefont {Gerland},\ and\ \citenamefont {Sutherland}}]{sutherland_2009}%
  \BibitemOpen
  \bibfield  {author} {\bibinfo {author} {\bibfnamefont {M.~W.}\ \bibnamefont {Powner}}, \bibinfo {author} {\bibfnamefont {B.}~\bibnamefont {Gerland}},\ and\ \bibinfo {author} {\bibfnamefont {J.~D.}\ \bibnamefont {Sutherland}},\ }\bibfield  {title} {\bibinfo {title} {Synthesis of activated pyrimidine ribonucleotides in prebiotically plausible conditions},\ }\href@noop {} {\bibfield  {journal} {\bibinfo  {journal} {Nature}\ }\textbf {\bibinfo {volume} {459}},\ \bibinfo {pages} {239} (\bibinfo {year} {2009})}\BibitemShut {NoStop}%
\bibitem [{\citenamefont {Patel}\ \emph {et~al.}(2015)\citenamefont {Patel}, \citenamefont {Percivalle}, \citenamefont {Ritson}, \citenamefont {Duffy},\ and\ \citenamefont {Sutherland}}]{patel_2015}%
  \BibitemOpen
  \bibfield  {author} {\bibinfo {author} {\bibfnamefont {B.~H.}\ \bibnamefont {Patel}}, \bibinfo {author} {\bibfnamefont {C.}~\bibnamefont {Percivalle}}, \bibinfo {author} {\bibfnamefont {D.~J.}\ \bibnamefont {Ritson}}, \bibinfo {author} {\bibfnamefont {C.~D.}\ \bibnamefont {Duffy}},\ and\ \bibinfo {author} {\bibfnamefont {J.~D.}\ \bibnamefont {Sutherland}},\ }\bibfield  {title} {\bibinfo {title} {Common origins of {RNA}, protein and lipid precursors in a cyanosulfidic protometabolism},\ }\href@noop {} {\bibfield  {journal} {\bibinfo  {journal} {Nature Chemistry}\ }\textbf {\bibinfo {volume} {7}},\ \bibinfo {pages} {301} (\bibinfo {year} {2015})}\BibitemShut {NoStop}%
\bibitem [{\citenamefont {Fox}\ and\ \citenamefont {Harada}(1958)}]{fox1958thermal}%
  \BibitemOpen
  \bibfield  {author} {\bibinfo {author} {\bibfnamefont {S.~W.}\ \bibnamefont {Fox}}\ and\ \bibinfo {author} {\bibfnamefont {K.}~\bibnamefont {Harada}},\ }\bibfield  {title} {\bibinfo {title} {Thermal copolymerization of amino acids to a product resembling protein},\ }\href@noop {} {\bibfield  {journal} {\bibinfo  {journal} {Science}\ }\textbf {\bibinfo {volume} {128}},\ \bibinfo {pages} {1214} (\bibinfo {year} {1958})}\BibitemShut {NoStop}%
\bibitem [{\citenamefont {Fox}\ \emph {et~al.}(1959)\citenamefont {Fox}, \citenamefont {Harada},\ and\ \citenamefont {Kendrick}}]{fox1959production}%
  \BibitemOpen
  \bibfield  {author} {\bibinfo {author} {\bibfnamefont {S.~W.}\ \bibnamefont {Fox}}, \bibinfo {author} {\bibfnamefont {K.}~\bibnamefont {Harada}},\ and\ \bibinfo {author} {\bibfnamefont {J.}~\bibnamefont {Kendrick}},\ }\bibfield  {title} {\bibinfo {title} {Production of spherules from synthetic proteinoid and hot water},\ }\href@noop {} {\bibfield  {journal} {\bibinfo  {journal} {Science}\ }\textbf {\bibinfo {volume} {129}},\ \bibinfo {pages} {1221} (\bibinfo {year} {1959})}\BibitemShut {NoStop}%
\bibitem [{\citenamefont {Luisi}\ and\ \citenamefont {Varela}(1989)}]{luisi1989self}%
  \BibitemOpen
  \bibfield  {author} {\bibinfo {author} {\bibfnamefont {P.~L.}\ \bibnamefont {Luisi}}\ and\ \bibinfo {author} {\bibfnamefont {F.~J.}\ \bibnamefont {Varela}},\ }\bibfield  {title} {\bibinfo {title} {Self-replicating micelles— a chemical version of a minimal autopoietic system},\ }\href@noop {} {\bibfield  {journal} {\bibinfo  {journal} {Origins of Life and Evolution of the Biosphere}\ }\textbf {\bibinfo {volume} {19}},\ \bibinfo {pages} {633} (\bibinfo {year} {1989})}\BibitemShut {NoStop}%
\bibitem [{\citenamefont {Szostak}\ \emph {et~al.}(2001)\citenamefont {Szostak}, \citenamefont {Bartel},\ and\ \citenamefont {Luisi}}]{szostak_2001}%
  \BibitemOpen
  \bibfield  {author} {\bibinfo {author} {\bibfnamefont {J.~W.}\ \bibnamefont {Szostak}}, \bibinfo {author} {\bibfnamefont {D.~P.}\ \bibnamefont {Bartel}},\ and\ \bibinfo {author} {\bibfnamefont {P.~L.}\ \bibnamefont {Luisi}},\ }\bibfield  {title} {\bibinfo {title} {Synthesizing life},\ }\href@noop {} {\bibfield  {journal} {\bibinfo  {journal} {Nature}\ }\textbf {\bibinfo {volume} {409}},\ \bibinfo {pages} {387} (\bibinfo {year} {2001})}\BibitemShut {NoStop}%
\bibitem [{\citenamefont {Hanczyc}\ \emph {et~al.}(2003)\citenamefont {Hanczyc}, \citenamefont {Fujikawa},\ and\ \citenamefont {Szostak}}]{hanczyc2003experimental}%
  \BibitemOpen
  \bibfield  {author} {\bibinfo {author} {\bibfnamefont {M.~M.}\ \bibnamefont {Hanczyc}}, \bibinfo {author} {\bibfnamefont {S.~M.}\ \bibnamefont {Fujikawa}},\ and\ \bibinfo {author} {\bibfnamefont {J.~W.}\ \bibnamefont {Szostak}},\ }\bibfield  {title} {\bibinfo {title} {Experimental models of primitive cellular compartments: encapsulation, growth, and division},\ }\href@noop {} {\bibfield  {journal} {\bibinfo  {journal} {Science}\ }\textbf {\bibinfo {volume} {302}},\ \bibinfo {pages} {618} (\bibinfo {year} {2003})}\BibitemShut {NoStop}%
\bibitem [{\citenamefont {Noireaux}\ and\ \citenamefont {Libchaber}(2004)}]{Noireaux2004}%
  \BibitemOpen
  \bibfield  {author} {\bibinfo {author} {\bibfnamefont {V.}~\bibnamefont {Noireaux}}\ and\ \bibinfo {author} {\bibfnamefont {A.}~\bibnamefont {Libchaber}},\ }\bibfield  {title} {\bibinfo {title} {{A vesicle bioreactor as a step toward an artificial cell assembly.}},\ }\href@noop {} {\bibfield  {journal} {\bibinfo  {journal} {Proceedings of the National Academy of Sciences}\ }\textbf {\bibinfo {volume} {101}},\ \bibinfo {pages} {17669} (\bibinfo {year} {2004})}\BibitemShut {NoStop}%
\bibitem [{\citenamefont {Mansy}\ \emph {et~al.}(2008)\citenamefont {Mansy}, \citenamefont {Schrum}, \citenamefont {Krishnamurthy}, \citenamefont {Tobe}, \citenamefont {Treco},\ and\ \citenamefont {Szostak}}]{mansy_2008}%
  \BibitemOpen
  \bibfield  {author} {\bibinfo {author} {\bibfnamefont {S.~S.}\ \bibnamefont {Mansy}}, \bibinfo {author} {\bibfnamefont {J.~P.}\ \bibnamefont {Schrum}}, \bibinfo {author} {\bibfnamefont {M.}~\bibnamefont {Krishnamurthy}}, \bibinfo {author} {\bibfnamefont {S.}~\bibnamefont {Tobe}}, \bibinfo {author} {\bibfnamefont {D.~A.}\ \bibnamefont {Treco}},\ and\ \bibinfo {author} {\bibfnamefont {J.~W.}\ \bibnamefont {Szostak}},\ }\bibfield  {title} {\bibinfo {title} {Template-directed synthesis of a genetic polymer in a model protocell},\ }\href@noop {} {\bibfield  {journal} {\bibinfo  {journal} {Nature}\ }\textbf {\bibinfo {volume} {454}},\ \bibinfo {pages} {122} (\bibinfo {year} {2008})}\BibitemShut {NoStop}%
\bibitem [{\citenamefont {Schwille}\ \emph {et~al.}(2018)\citenamefont {Schwille}, \citenamefont {Spatz}, \citenamefont {Landfester}, \citenamefont {Bodenschatz}, \citenamefont {Herminghaus}, \citenamefont {Sourjik}, \citenamefont {Erb}, \citenamefont {Bastiaens}, \citenamefont {Lipowsky}, \citenamefont {Hyman} \emph {et~al.}}]{schwille2018maxsynbio}%
  \BibitemOpen
  \bibfield  {author} {\bibinfo {author} {\bibfnamefont {P.}~\bibnamefont {Schwille}}, \bibinfo {author} {\bibfnamefont {J.}~\bibnamefont {Spatz}}, \bibinfo {author} {\bibfnamefont {K.}~\bibnamefont {Landfester}}, \bibinfo {author} {\bibfnamefont {E.}~\bibnamefont {Bodenschatz}}, \bibinfo {author} {\bibfnamefont {S.}~\bibnamefont {Herminghaus}}, \bibinfo {author} {\bibfnamefont {V.}~\bibnamefont {Sourjik}}, \bibinfo {author} {\bibfnamefont {T.~J.}\ \bibnamefont {Erb}}, \bibinfo {author} {\bibfnamefont {P.}~\bibnamefont {Bastiaens}}, \bibinfo {author} {\bibfnamefont {R.}~\bibnamefont {Lipowsky}}, \bibinfo {author} {\bibfnamefont {A.}~\bibnamefont {Hyman}}, \emph {et~al.},\ }\bibfield  {title} {\bibinfo {title} {Maxsynbio: avenues towards creating cells from the bottom up},\ }\href@noop {} {\bibfield  {journal} {\bibinfo  {journal} {Angewandte Chemie International Edition}\ }\textbf {\bibinfo {volume} {57}},\ \bibinfo {pages} {13382} (\bibinfo {year} {2018})}\BibitemShut {NoStop}%
\bibitem [{\citenamefont {Morrow}\ \emph {et~al.}(2019)\citenamefont {Morrow}, \citenamefont {Colomer},\ and\ \citenamefont {Fletcher}}]{morrow2019chemically}%
  \BibitemOpen
  \bibfield  {author} {\bibinfo {author} {\bibfnamefont {S.~M.}\ \bibnamefont {Morrow}}, \bibinfo {author} {\bibfnamefont {I.}~\bibnamefont {Colomer}},\ and\ \bibinfo {author} {\bibfnamefont {S.~P.}\ \bibnamefont {Fletcher}},\ }\bibfield  {title} {\bibinfo {title} {A chemically fuelled self-replicator},\ }\href@noop {} {\bibfield  {journal} {\bibinfo  {journal} {Nature Communications}\ }\textbf {\bibinfo {volume} {10}},\ \bibinfo {pages} {1011} (\bibinfo {year} {2019})}\BibitemShut {NoStop}%
\bibitem [{\citenamefont {Cho}\ \emph {et~al.}(2024)\citenamefont {Cho}, \citenamefont {An}, \citenamefont {Lai}, \citenamefont {Fracassi}, \citenamefont {Brea}, \citenamefont {Chen},\ and\ \citenamefont {Devaraj}}]{cho_2024}%
  \BibitemOpen
  \bibfield  {author} {\bibinfo {author} {\bibfnamefont {C.~J.}\ \bibnamefont {Cho}}, \bibinfo {author} {\bibfnamefont {T.}~\bibnamefont {An}}, \bibinfo {author} {\bibfnamefont {Y.-C.}\ \bibnamefont {Lai}}, \bibinfo {author} {\bibfnamefont {A.}~\bibnamefont {Fracassi}}, \bibinfo {author} {\bibfnamefont {R.~J.}\ \bibnamefont {Brea}}, \bibinfo {author} {\bibfnamefont {I.~A.}\ \bibnamefont {Chen}},\ and\ \bibinfo {author} {\bibfnamefont {N.~K.}\ \bibnamefont {Devaraj}},\ }\bibfield  {title} {\bibinfo {title} {Protocells by spontaneous reaction of cysteine with short-chain thioesters},\ }\href@noop {} {\bibfield  {journal} {\bibinfo  {journal} {Nature Chemistry}\ }\textbf {\bibinfo {volume} {17}},\ \bibinfo {pages} {148} (\bibinfo {year} {2024})}\BibitemShut {NoStop}%
\bibitem [{\citenamefont {Katla}\ \emph {et~al.}(2025)\citenamefont {Katla}, \citenamefont {Lin},\ and\ \citenamefont {P{\'e}rez-Mercader}}]{katla2025self}%
  \BibitemOpen
  \bibfield  {author} {\bibinfo {author} {\bibfnamefont {S.~K.}\ \bibnamefont {Katla}}, \bibinfo {author} {\bibfnamefont {C.}~\bibnamefont {Lin}},\ and\ \bibinfo {author} {\bibfnamefont {J.}~\bibnamefont {P{\'e}rez-Mercader}},\ }\bibfield  {title} {\bibinfo {title} {Self-reproduction as an autonomous process of growth and reorganization in fully abiotic, artificial and synthetic cells},\ }\href@noop {} {\bibfield  {journal} {\bibinfo  {journal} {Proceedings of the National Academy of Sciences}\ }\textbf {\bibinfo {volume} {122}},\ \bibinfo {pages} {e2412514122} (\bibinfo {year} {2025})}\BibitemShut {NoStop}%
\bibitem [{\citenamefont {Wenisch}\ \emph {et~al.}(2025)\citenamefont {Wenisch}, \citenamefont {Li}, \citenamefont {Braun}, \citenamefont {Eylert}, \citenamefont {Sp{\"a}th}, \citenamefont {Poprawa}, \citenamefont {Rieger}, \citenamefont {Synatschke}, \citenamefont {Niederholtmeyer},\ and\ \citenamefont {Boekhoven}}]{wenisch2025toward}%
  \BibitemOpen
  \bibfield  {author} {\bibinfo {author} {\bibfnamefont {M.}~\bibnamefont {Wenisch}}, \bibinfo {author} {\bibfnamefont {Y.}~\bibnamefont {Li}}, \bibinfo {author} {\bibfnamefont {M.~G.}\ \bibnamefont {Braun}}, \bibinfo {author} {\bibfnamefont {L.}~\bibnamefont {Eylert}}, \bibinfo {author} {\bibfnamefont {F.}~\bibnamefont {Sp{\"a}th}}, \bibinfo {author} {\bibfnamefont {S.~M.}\ \bibnamefont {Poprawa}}, \bibinfo {author} {\bibfnamefont {B.}~\bibnamefont {Rieger}}, \bibinfo {author} {\bibfnamefont {C.~V.}\ \bibnamefont {Synatschke}}, \bibinfo {author} {\bibfnamefont {H.}~\bibnamefont {Niederholtmeyer}},\ and\ \bibinfo {author} {\bibfnamefont {J.}~\bibnamefont {Boekhoven}},\ }\bibfield  {title} {\bibinfo {title} {Toward synthetic life—emergence, growth, creation of offspring, decay, and rescue of fuel-dependent synthetic cells},\ }\href@noop {} {\bibfield  {journal} {\bibinfo  {journal} {Chem}\ } (\bibinfo {year} {2025})}\BibitemShut {NoStop}%
\bibitem [{\citenamefont {Prigogine}\ and\ \citenamefont {Nicolis}(1985)}]{prigogine1985self}%
  \BibitemOpen
  \bibfield  {author} {\bibinfo {author} {\bibfnamefont {I.}~\bibnamefont {Prigogine}}\ and\ \bibinfo {author} {\bibfnamefont {G.}~\bibnamefont {Nicolis}},\ }\bibfield  {title} {\bibinfo {title} {Self-organisation in nonequilibrium systems: towards a dynamics of complexity},\ }in\ \href@noop {} {\emph {\bibinfo {booktitle} {Bifurcation analysis}}}\ (\bibinfo  {publisher} {Springer},\ \bibinfo {year} {1985})\ pp.\ \bibinfo {pages} {3--12}\BibitemShut {NoStop}%
\bibitem [{\citenamefont {G{\'a}nti}(2003{\natexlab{b}})}]{ganti2003chemoton}%
  \BibitemOpen
  \bibfield  {author} {\bibinfo {author} {\bibfnamefont {T.}~\bibnamefont {G{\'a}nti}},\ }\href@noop {} {\emph {\bibinfo {title} {Chemoton theory: theory of living systems}}}\ (\bibinfo  {publisher} {Springer Science \& Business Media},\ \bibinfo {year} {2003})\BibitemShut {NoStop}%
\bibitem [{\citenamefont {Benner}\ \emph {et~al.}(2018)\citenamefont {Benner}, \citenamefont {Kim},\ and\ \citenamefont {Biondi}}]{benner2018mineral}%
  \BibitemOpen
  \bibfield  {author} {\bibinfo {author} {\bibfnamefont {S.~A.}\ \bibnamefont {Benner}}, \bibinfo {author} {\bibfnamefont {H.-J.}\ \bibnamefont {Kim}},\ and\ \bibinfo {author} {\bibfnamefont {E.}~\bibnamefont {Biondi}},\ }\bibfield  {title} {\bibinfo {title} {Mineral-organic interactions in prebiotic synthesis: The discontinuous synthesis model for the formation of {RNA} in naturally complex geological environments},\ }in\ \href@noop {} {\emph {\bibinfo {booktitle} {Prebiotic Chemistry and Chemical Evolution of Nucleic Acids}}}\ (\bibinfo  {publisher} {Springer},\ \bibinfo {year} {2018})\ pp.\ \bibinfo {pages} {31--83}\BibitemShut {NoStop}%
\bibitem [{\citenamefont {Bahadur}\ and\ \citenamefont {Ranganayaki}(1964)}]{bahadur1964synthesis1}%
  \BibitemOpen
  \bibfield  {author} {\bibinfo {author} {\bibfnamefont {K.}~\bibnamefont {Bahadur}}\ and\ \bibinfo {author} {\bibfnamefont {S.}~\bibnamefont {Ranganayaki}},\ }\bibfield  {title} {\bibinfo {title} {Synthesis of {J}eewanu, the units capable of growth, multiplication and metabolic activity. {I}. {P}reparation of units capable of growth and division and having metabolic activity},\ }\href@noop {} {\bibfield  {journal} {\bibinfo  {journal} {Zentr. Bakteriol. Parasitenk}\ }\textbf {\bibinfo {volume} {117}},\ \bibinfo {pages} {567} (\bibinfo {year} {1964})}\BibitemShut {NoStop}%
\bibitem [{\citenamefont {Bahadur}\ \emph {et~al.}(1964)\citenamefont {Bahadur} \emph {et~al.}}]{bahadur1964synthesis2}%
  \BibitemOpen
  \bibfield  {author} {\bibinfo {author} {\bibfnamefont {K.}~\bibnamefont {Bahadur}} \emph {et~al.},\ }\bibfield  {title} {\bibinfo {title} {Synthesis of {J}eewanu, the units capable of growth, multiplication and metabolic activity. {II}. {P}hotochemical preparation of growing and multiplying units with metabolic activities},\ }\href@noop {} {\bibfield  {journal} {\bibinfo  {journal} {Zentr. Bakteriol. Parasitenk}\ }\textbf {\bibinfo {volume} {117}},\ \bibinfo {pages} {575} (\bibinfo {year} {1964})}\BibitemShut {NoStop}%
\bibitem [{\citenamefont {Bahadur}(1964)}]{bahadur1964synthesis3}%
  \BibitemOpen
  \bibfield  {author} {\bibinfo {author} {\bibfnamefont {K.}~\bibnamefont {Bahadur}},\ }\bibfield  {title} {\bibinfo {title} {Synthesis of {J}eewanu, the units capable of growth, multiplication and metabolic activity. {III}. {P}reparation of microspheres capable of growth and division by budding and having metabolic activity with peptides prepared thermally},\ }\href@noop {} {\bibfield  {journal} {\bibinfo  {journal} {Zentr. Bakteriol. Parasitenk}\ }\textbf {\bibinfo {volume} {117}},\ \bibinfo {pages} {585} (\bibinfo {year} {1964})}\BibitemShut {NoStop}%
\bibitem [{\citenamefont {Liu}\ \emph {et~al.}(2003)\citenamefont {Liu}, \citenamefont {Diemann}, \citenamefont {Li}, \citenamefont {Dress},\ and\ \citenamefont {M{\"u}ller}}]{liu2003self}%
  \BibitemOpen
  \bibfield  {author} {\bibinfo {author} {\bibfnamefont {T.}~\bibnamefont {Liu}}, \bibinfo {author} {\bibfnamefont {E.}~\bibnamefont {Diemann}}, \bibinfo {author} {\bibfnamefont {H.}~\bibnamefont {Li}}, \bibinfo {author} {\bibfnamefont {A.~W.}\ \bibnamefont {Dress}},\ and\ \bibinfo {author} {\bibfnamefont {A.}~\bibnamefont {M{\"u}ller}},\ }\bibfield  {title} {\bibinfo {title} {Self-assembly in aqueous solution of wheel-shaped {M}o$_{154}$ oxide clusters into vesicles},\ }\href@noop {} {\bibfield  {journal} {\bibinfo  {journal} {Nature}\ }\textbf {\bibinfo {volume} {426}},\ \bibinfo {pages} {59} (\bibinfo {year} {2003})}\BibitemShut {NoStop}%
\bibitem [{\citenamefont {Miras}\ \emph {et~al.}(2020)\citenamefont {Miras}, \citenamefont {Mathis}, \citenamefont {Xuan}, \citenamefont {Long}, \citenamefont {Pow},\ and\ \citenamefont {Cronin}}]{miras2020spontaneous}%
  \BibitemOpen
  \bibfield  {author} {\bibinfo {author} {\bibfnamefont {H.~N.}\ \bibnamefont {Miras}}, \bibinfo {author} {\bibfnamefont {C.}~\bibnamefont {Mathis}}, \bibinfo {author} {\bibfnamefont {W.}~\bibnamefont {Xuan}}, \bibinfo {author} {\bibfnamefont {D.-L.}\ \bibnamefont {Long}}, \bibinfo {author} {\bibfnamefont {R.}~\bibnamefont {Pow}},\ and\ \bibinfo {author} {\bibfnamefont {L.}~\bibnamefont {Cronin}},\ }\bibfield  {title} {\bibinfo {title} {Spontaneous formation of autocatalytic sets with self-replicating inorganic metal oxide clusters},\ }\href@noop {} {\bibfield  {journal} {\bibinfo  {journal} {Proceedings of the National Academy of Sciences}\ }\textbf {\bibinfo {volume} {117}},\ \bibinfo {pages} {10699} (\bibinfo {year} {2020})}\BibitemShut {NoStop}%
\bibitem [{\citenamefont {Cleaves~II}(2008)}]{cleaves2008prebiotic}%
  \BibitemOpen
  \bibfield  {author} {\bibinfo {author} {\bibfnamefont {H.~J.}\ \bibnamefont {Cleaves~II}},\ }\bibfield  {title} {\bibinfo {title} {The prebiotic geochemistry of formaldehyde},\ }\href@noop {} {\bibfield  {journal} {\bibinfo  {journal} {Precambrian Research}\ }\textbf {\bibinfo {volume} {164}},\ \bibinfo {pages} {111} (\bibinfo {year} {2008})}\BibitemShut {NoStop}%
\bibitem [{\citenamefont {Todd}(2022)}]{todd2022sources}%
  \BibitemOpen
  \bibfield  {author} {\bibinfo {author} {\bibfnamefont {Z.~R.}\ \bibnamefont {Todd}},\ }\bibfield  {title} {\bibinfo {title} {Sources of nitrogen-, sulfur-, and phosphorus-containing feedstocks for prebiotic chemistry in the planetary environment},\ }\href@noop {} {\bibfield  {journal} {\bibinfo  {journal} {Life}\ }\textbf {\bibinfo {volume} {12}},\ \bibinfo {pages} {1268} (\bibinfo {year} {2022})}\BibitemShut {NoStop}%
\bibitem [{\citenamefont {M{\"u}ller}\ and\ \citenamefont {Serain}(2000)}]{muller2000soluble}%
  \BibitemOpen
  \bibfield  {author} {\bibinfo {author} {\bibfnamefont {A.}~\bibnamefont {M{\"u}ller}}\ and\ \bibinfo {author} {\bibfnamefont {C.}~\bibnamefont {Serain}},\ }\bibfield  {title} {\bibinfo {title} {Soluble molybdenum blues “des pudels kern”},\ }\href@noop {} {\bibfield  {journal} {\bibinfo  {journal} {Accounts of Chemical Research}\ }\textbf {\bibinfo {volume} {33}},\ \bibinfo {pages} {2} (\bibinfo {year} {2000})}\BibitemShut {NoStop}%
\bibitem [{\citenamefont {Krevelen}(1950)}]{krevelen1950graphical}%
  \BibitemOpen
  \bibfield  {author} {\bibinfo {author} {\bibfnamefont {V.}~\bibnamefont {Krevelen}},\ }\bibfield  {title} {\bibinfo {title} {Graphical-statistical method for the study of structure and reaction processes of coal.},\ }\href@noop {} {\bibfield  {journal} {\bibinfo  {journal} {Fuel}\ }\textbf {\bibinfo {volume} {29}},\ \bibinfo {pages} {269} (\bibinfo {year} {1950})}\BibitemShut {NoStop}%
\bibitem [{\citenamefont {Kim}\ \emph {et~al.}(2003)\citenamefont {Kim}, \citenamefont {Kramer},\ and\ \citenamefont {Hatcher}}]{kim2003graphical}%
  \BibitemOpen
  \bibfield  {author} {\bibinfo {author} {\bibfnamefont {S.}~\bibnamefont {Kim}}, \bibinfo {author} {\bibfnamefont {R.~W.}\ \bibnamefont {Kramer}},\ and\ \bibinfo {author} {\bibfnamefont {P.~G.}\ \bibnamefont {Hatcher}},\ }\bibfield  {title} {\bibinfo {title} {Graphical method for analysis of ultrahigh-resolution broadband mass spectra of natural organic matter, the van {K}revelen diagram},\ }\href@noop {} {\bibfield  {journal} {\bibinfo  {journal} {Analytical Chemistry}\ }\textbf {\bibinfo {volume} {75}},\ \bibinfo {pages} {5336} (\bibinfo {year} {2003})}\BibitemShut {NoStop}%
\bibitem [{\citenamefont {Rivas-Ubach}\ \emph {et~al.}(2018)\citenamefont {Rivas-Ubach}, \citenamefont {Liu}, \citenamefont {Bianchi}, \citenamefont {Tolic}, \citenamefont {Jansson},\ and\ \citenamefont {Pasa-Tolic}}]{rivas2018moving}%
  \BibitemOpen
  \bibfield  {author} {\bibinfo {author} {\bibfnamefont {A.}~\bibnamefont {Rivas-Ubach}}, \bibinfo {author} {\bibfnamefont {Y.}~\bibnamefont {Liu}}, \bibinfo {author} {\bibfnamefont {T.~S.}\ \bibnamefont {Bianchi}}, \bibinfo {author} {\bibfnamefont {N.}~\bibnamefont {Tolic}}, \bibinfo {author} {\bibfnamefont {C.}~\bibnamefont {Jansson}},\ and\ \bibinfo {author} {\bibfnamefont {L.}~\bibnamefont {Pasa-Tolic}},\ }\bibfield  {title} {\bibinfo {title} {Moving beyond the van krevelen diagram: A new stoichiometric approach for compound classification in organisms},\ }\href@noop {} {\bibfield  {journal} {\bibinfo  {journal} {Analytical Chemistry}\ }\textbf {\bibinfo {volume} {90}},\ \bibinfo {pages} {6152} (\bibinfo {year} {2018})}\BibitemShut {NoStop}%
\bibitem [{\citenamefont {Saji}\ and\ \citenamefont {Lee}(2012)}]{saji2012molybdenum}%
  \BibitemOpen
  \bibfield  {author} {\bibinfo {author} {\bibfnamefont {V.~S.}\ \bibnamefont {Saji}}\ and\ \bibinfo {author} {\bibfnamefont {C.-W.}\ \bibnamefont {Lee}},\ }\bibfield  {title} {\bibinfo {title} {Molybdenum, molybdenum oxides, and their electrochemistry},\ }\href@noop {} {\bibfield  {journal} {\bibinfo  {journal} {ChemSusChem}\ }\textbf {\bibinfo {volume} {5}},\ \bibinfo {pages} {1146} (\bibinfo {year} {2012})}\BibitemShut {NoStop}%
\bibitem [{\citenamefont {Morowitz}\ and\ \citenamefont {Smith}(2007)}]{morowitz2007energy}%
  \BibitemOpen
  \bibfield  {author} {\bibinfo {author} {\bibfnamefont {H.}~\bibnamefont {Morowitz}}\ and\ \bibinfo {author} {\bibfnamefont {E.}~\bibnamefont {Smith}},\ }\bibfield  {title} {\bibinfo {title} {Energy flow and the organization of life},\ }\href@noop {} {\bibfield  {journal} {\bibinfo  {journal} {Complexity}\ }\textbf {\bibinfo {volume} {13}},\ \bibinfo {pages} {51} (\bibinfo {year} {2007})}\BibitemShut {NoStop}%
\bibitem [{\citenamefont {Muchowska}\ \emph {et~al.}(2020)\citenamefont {Muchowska}, \citenamefont {Varma},\ and\ \citenamefont {Moran}}]{muchowska2020nonenzymatic}%
  \BibitemOpen
  \bibfield  {author} {\bibinfo {author} {\bibfnamefont {K.~B.}\ \bibnamefont {Muchowska}}, \bibinfo {author} {\bibfnamefont {S.~J.}\ \bibnamefont {Varma}},\ and\ \bibinfo {author} {\bibfnamefont {J.}~\bibnamefont {Moran}},\ }\bibfield  {title} {\bibinfo {title} {Nonenzymatic metabolic reactions and life’s origins},\ }\href@noop {} {\bibfield  {journal} {\bibinfo  {journal} {Chemical Reviews}\ }\textbf {\bibinfo {volume} {120}},\ \bibinfo {pages} {7708} (\bibinfo {year} {2020})}\BibitemShut {NoStop}%
\bibitem [{\citenamefont {Helz}\ \emph {et~al.}(2004)\citenamefont {Helz}, \citenamefont {Vorlicek},\ and\ \citenamefont {Kahn}}]{helz2004molybdenum}%
  \BibitemOpen
  \bibfield  {author} {\bibinfo {author} {\bibfnamefont {G.~R.}\ \bibnamefont {Helz}}, \bibinfo {author} {\bibfnamefont {T.~P.}\ \bibnamefont {Vorlicek}},\ and\ \bibinfo {author} {\bibfnamefont {M.~D.}\ \bibnamefont {Kahn}},\ }\bibfield  {title} {\bibinfo {title} {Molybdenum scavenging by iron monosulfide},\ }\href@noop {} {\bibfield  {journal} {\bibinfo  {journal} {Environmental Science \& Technology}\ }\textbf {\bibinfo {volume} {38}},\ \bibinfo {pages} {4263} (\bibinfo {year} {2004})}\BibitemShut {NoStop}%
\bibitem [{\citenamefont {Inaba}(2018)}]{inaba2018primary}%
  \BibitemOpen
  \bibfield  {author} {\bibinfo {author} {\bibfnamefont {S.}~\bibnamefont {Inaba}},\ }\bibfield  {title} {\bibinfo {title} {Primary formation path of formaldehyde in hydrothermal vents},\ }\href@noop {} {\bibfield  {journal} {\bibinfo  {journal} {Origins of Life and Evolution of Biospheres}\ }\textbf {\bibinfo {volume} {48}},\ \bibinfo {pages} {1} (\bibinfo {year} {2018})}\BibitemShut {NoStop}%
\bibitem [{\citenamefont {Evans}\ \emph {et~al.}(2023)\citenamefont {Evans}, \citenamefont {Coogan}, \citenamefont {Ka{\c{c}}ar},\ and\ \citenamefont {Seyfried}}]{evans2023molybdenum}%
  \BibitemOpen
  \bibfield  {author} {\bibinfo {author} {\bibfnamefont {G.~N.}\ \bibnamefont {Evans}}, \bibinfo {author} {\bibfnamefont {L.~A.}\ \bibnamefont {Coogan}}, \bibinfo {author} {\bibfnamefont {B.}~\bibnamefont {Ka{\c{c}}ar}},\ and\ \bibinfo {author} {\bibfnamefont {W.~E.}\ \bibnamefont {Seyfried}},\ }\bibfield  {title} {\bibinfo {title} {Molybdenum in basalt-hosted seafloor hydrothermal systems: experimental, theoretical, and field sampling approaches},\ }\href@noop {} {\bibfield  {journal} {\bibinfo  {journal} {Geochimica et Cosmochimica Acta}\ }\textbf {\bibinfo {volume} {353}},\ \bibinfo {pages} {28} (\bibinfo {year} {2023})}\BibitemShut {NoStop}%
\bibitem [{\citenamefont {Takahagi}\ \emph {et~al.}(2023)\citenamefont {Takahagi}, \citenamefont {Okada}, \citenamefont {Matsui}, \citenamefont {Ono}, \citenamefont {Takai}, \citenamefont {Takahashi},\ and\ \citenamefont {Kitadai}}]{takahagi2023extreme}%
  \BibitemOpen
  \bibfield  {author} {\bibinfo {author} {\bibfnamefont {W.}~\bibnamefont {Takahagi}}, \bibinfo {author} {\bibfnamefont {S.}~\bibnamefont {Okada}}, \bibinfo {author} {\bibfnamefont {Y.}~\bibnamefont {Matsui}}, \bibinfo {author} {\bibfnamefont {S.}~\bibnamefont {Ono}}, \bibinfo {author} {\bibfnamefont {K.}~\bibnamefont {Takai}}, \bibinfo {author} {\bibfnamefont {Y.}~\bibnamefont {Takahashi}},\ and\ \bibinfo {author} {\bibfnamefont {N.}~\bibnamefont {Kitadai}},\ }\bibfield  {title} {\bibinfo {title} {Extreme accumulation of ammonia on electroreduced mackinawite: An abiotic ammonia storage mechanism in early ocean hydrothermal systems},\ }\href@noop {} {\bibfield  {journal} {\bibinfo  {journal} {Proceedings of the National Academy of Sciences}\ }\textbf {\bibinfo {volume} {120}},\ \bibinfo {pages} {e2303302120} (\bibinfo {year} {2023})}\BibitemShut {NoStop}%
\bibitem [{\citenamefont {Randolph-Flagg}\ \emph {et~al.}(2023)\citenamefont {Randolph-Flagg}, \citenamefont {Ely}, \citenamefont {Som}, \citenamefont {Shock}, \citenamefont {German},\ and\ \citenamefont {Hoehler}}]{randolph2023phosphate}%
  \BibitemOpen
  \bibfield  {author} {\bibinfo {author} {\bibfnamefont {N.~G.}\ \bibnamefont {Randolph-Flagg}}, \bibinfo {author} {\bibfnamefont {T.}~\bibnamefont {Ely}}, \bibinfo {author} {\bibfnamefont {S.~M.}\ \bibnamefont {Som}}, \bibinfo {author} {\bibfnamefont {E.~L.}\ \bibnamefont {Shock}}, \bibinfo {author} {\bibfnamefont {C.~R.}\ \bibnamefont {German}},\ and\ \bibinfo {author} {\bibfnamefont {T.~M.}\ \bibnamefont {Hoehler}},\ }\bibfield  {title} {\bibinfo {title} {Phosphate availability and implications for life on ocean worlds},\ }\href@noop {} {\bibfield  {journal} {\bibinfo  {journal} {Nature Communications}\ }\textbf {\bibinfo {volume} {14}},\ \bibinfo {pages} {2388} (\bibinfo {year} {2023})}\BibitemShut {NoStop}%
\bibitem [{\citenamefont {Mitchell}(1966)}]{classics1966chemiosmotic}%
  \BibitemOpen
  \bibfield  {author} {\bibinfo {author} {\bibfnamefont {P.}~\bibnamefont {Mitchell}},\ }\bibfield  {title} {\bibinfo {title} {Chemiosmotic coupling in oxidative and photosynthetic phosphorylation},\ }\href@noop {} {\bibfield  {journal} {\bibinfo  {journal} {Biol. Rev. Cambridge Phil Soc}\ }\textbf {\bibinfo {volume} {41}},\ \bibinfo {pages} {445} (\bibinfo {year} {1966})}\BibitemShut {NoStop}%
\bibitem [{\citenamefont {Kauffman}(1971)}]{kauffman1971cellular}%
  \BibitemOpen
  \bibfield  {author} {\bibinfo {author} {\bibfnamefont {S.~A.}\ \bibnamefont {Kauffman}},\ }\bibfield  {title} {\bibinfo {title} {Cellular homeostasis, epigenesis and replication in randomly aggregated macromolecular systems},\ }\href@noop {} {\bibfield  {journal} {\bibinfo  {journal} {Journal of Cybernetics}\ }\textbf {\bibinfo {volume} {1}},\ \bibinfo {pages} {71} (\bibinfo {year} {1971})}\BibitemShut {NoStop}%
\bibitem [{\citenamefont {Corliss}(1990)}]{corliss1990hot}%
  \BibitemOpen
  \bibfield  {author} {\bibinfo {author} {\bibfnamefont {J.~B.}\ \bibnamefont {Corliss}},\ }\bibfield  {title} {\bibinfo {title} {Hot springs and the origin of life},\ }\href@noop {} {\bibfield  {journal} {\bibinfo  {journal} {Nature}\ }\textbf {\bibinfo {volume} {347}},\ \bibinfo {pages} {624} (\bibinfo {year} {1990})}\BibitemShut {NoStop}%
\bibitem [{\citenamefont {Lane}(2015)}]{lane2015vital}%
  \BibitemOpen
  \bibfield  {author} {\bibinfo {author} {\bibfnamefont {N.}~\bibnamefont {Lane}},\ }\href@noop {} {\emph {\bibinfo {title} {Vital Question: Energy, Evolution, and the Origins of Complex Life}}}\ (\bibinfo  {publisher} {WW Norton \& Company},\ \bibinfo {year} {2015})\BibitemShut {NoStop}%
\bibitem [{\citenamefont {Harrison}\ \emph {et~al.}(2023)\citenamefont {Harrison} \emph {et~al.}}]{harrison2023life}%
  \BibitemOpen
  \bibfield  {author} {\bibinfo {author} {\bibfnamefont {S.~A.}\ \bibnamefont {Harrison}} \emph {et~al.},\ }\bibfield  {title} {\bibinfo {title} {Life as a guide to its own origins},\ }\href@noop {} {\bibfield  {journal} {\bibinfo  {journal} {Annual Review of Ecology, Evolution, and Systematics}\ }\textbf {\bibinfo {volume} {54}},\ \bibinfo {pages} {327} (\bibinfo {year} {2023})}\BibitemShut {NoStop}%
\bibitem [{\citenamefont {Kauffman}(1995)}]{kauffman1995home}%
  \BibitemOpen
  \bibfield  {author} {\bibinfo {author} {\bibfnamefont {S.~A.}\ \bibnamefont {Kauffman}},\ }\href@noop {} {\emph {\bibinfo {title} {At home in the universe: The search for laws of self-organization and complexity}}}\ (\bibinfo  {publisher} {Oxford University Press, USA},\ \bibinfo {year} {1995})\BibitemShut {NoStop}%
\bibitem [{\citenamefont {Bissette}\ and\ \citenamefont {Fletcher}(2013)}]{bissette2013mechanisms}%
  \BibitemOpen
  \bibfield  {author} {\bibinfo {author} {\bibfnamefont {A.~J.}\ \bibnamefont {Bissette}}\ and\ \bibinfo {author} {\bibfnamefont {S.~P.}\ \bibnamefont {Fletcher}},\ }\bibfield  {title} {\bibinfo {title} {Mechanisms of autocatalysis},\ }\href@noop {} {\bibfield  {journal} {\bibinfo  {journal} {Angewandte Chemie International Edition}\ }\textbf {\bibinfo {volume} {52}},\ \bibinfo {pages} {12800} (\bibinfo {year} {2013})}\BibitemShut {NoStop}%
\bibitem [{\citenamefont {Davies}(1999)}]{davies1999fifth}%
  \BibitemOpen
  \bibfield  {author} {\bibinfo {author} {\bibfnamefont {P.}~\bibnamefont {Davies}},\ }\href@noop {} {\emph {\bibinfo {title} {The fifth miracle: The search for the origin and meaning of life}}}\ (\bibinfo  {publisher} {Simon and Schuster},\ \bibinfo {year} {1999})\BibitemShut {NoStop}%
\bibitem [{\citenamefont {Kauffman}(2000)}]{kauffman2000investigations}%
  \BibitemOpen
  \bibfield  {author} {\bibinfo {author} {\bibfnamefont {S.~A.}\ \bibnamefont {Kauffman}},\ }\href@noop {} {\emph {\bibinfo {title} {Investigations}}}\ (\bibinfo  {publisher} {Oxford University Press},\ \bibinfo {year} {2000})\BibitemShut {NoStop}%
\bibitem [{\citenamefont {Hurowitz}\ \emph {et~al.}(2025)\citenamefont {Hurowitz}, \citenamefont {Tice}, \citenamefont {Allwood}, \citenamefont {Cable}, \citenamefont {Hand}, \citenamefont {Murphy}, \citenamefont {Uckert}, \citenamefont {Bell~III}, \citenamefont {Bosak}, \citenamefont {Broz} \emph {et~al.}}]{hurowitz2025redox}%
  \BibitemOpen
  \bibfield  {author} {\bibinfo {author} {\bibfnamefont {J.~A.}\ \bibnamefont {Hurowitz}}, \bibinfo {author} {\bibfnamefont {M.}~\bibnamefont {Tice}}, \bibinfo {author} {\bibfnamefont {A.}~\bibnamefont {Allwood}}, \bibinfo {author} {\bibfnamefont {M.}~\bibnamefont {Cable}}, \bibinfo {author} {\bibfnamefont {K.}~\bibnamefont {Hand}}, \bibinfo {author} {\bibfnamefont {A.}~\bibnamefont {Murphy}}, \bibinfo {author} {\bibfnamefont {K.}~\bibnamefont {Uckert}}, \bibinfo {author} {\bibfnamefont {J.}~\bibnamefont {Bell~III}}, \bibinfo {author} {\bibfnamefont {T.}~\bibnamefont {Bosak}}, \bibinfo {author} {\bibfnamefont {A.}~\bibnamefont {Broz}}, \emph {et~al.},\ }\bibfield  {title} {\bibinfo {title} {Redox-driven mineral and organic associations in jezero crater, mars},\ }\href@noop {} {\bibfield  {journal} {\bibinfo  {journal} {Nature}\ }\textbf {\bibinfo {volume} {645}},\ \bibinfo {pages} {332} (\bibinfo {year} {2025})}\BibitemShut {NoStop}%
\bibitem [{\citenamefont {Schwieterman}\ \emph {et~al.}(2018)\citenamefont {Schwieterman}, \citenamefont {Kiang}, \citenamefont {Parenteau}, \citenamefont {Harman}, \citenamefont {DasSarma}, \citenamefont {Fisher}, \citenamefont {Arney}, \citenamefont {Hartnett}, \citenamefont {Reinhard}, \citenamefont {Olson} \emph {et~al.}}]{schwieterman2018exoplanet}%
  \BibitemOpen
  \bibfield  {author} {\bibinfo {author} {\bibfnamefont {E.~W.}\ \bibnamefont {Schwieterman}}, \bibinfo {author} {\bibfnamefont {N.~Y.}\ \bibnamefont {Kiang}}, \bibinfo {author} {\bibfnamefont {M.~N.}\ \bibnamefont {Parenteau}}, \bibinfo {author} {\bibfnamefont {C.~E.}\ \bibnamefont {Harman}}, \bibinfo {author} {\bibfnamefont {S.}~\bibnamefont {DasSarma}}, \bibinfo {author} {\bibfnamefont {T.~M.}\ \bibnamefont {Fisher}}, \bibinfo {author} {\bibfnamefont {G.~N.}\ \bibnamefont {Arney}}, \bibinfo {author} {\bibfnamefont {H.~E.}\ \bibnamefont {Hartnett}}, \bibinfo {author} {\bibfnamefont {C.~T.}\ \bibnamefont {Reinhard}}, \bibinfo {author} {\bibfnamefont {S.~L.}\ \bibnamefont {Olson}}, \emph {et~al.},\ }\bibfield  {title} {\bibinfo {title} {Exoplanet biosignatures: a review of remotely detectable signs of life},\ }\href@noop {} {\bibfield  {journal} {\bibinfo  {journal} {Astrobiology}\ }\textbf {\bibinfo {volume} {18}},\ \bibinfo {pages} {663} (\bibinfo {year} {2018})}\BibitemShut {NoStop}%
\bibitem [{\citenamefont {of~Sciences}\ \emph {et~al.}(2019)\citenamefont {of~Sciences}, \citenamefont {Medicine}, \citenamefont {on~Engineering}, \citenamefont {Sciences}, \citenamefont {Board},\ and\ \citenamefont {on~Astrobiology Science Strategy for the Search for Life in~the Universe}}]{national2019astrobiology}%
  \BibitemOpen
  \bibfield  {author} {\bibinfo {author} {\bibfnamefont {N.~A.}\ \bibnamefont {of~Sciences}}, \bibinfo {author} {\bibnamefont {Medicine}}, \bibinfo {author} {\bibfnamefont {D.}~\bibnamefont {on~Engineering}}, \bibinfo {author} {\bibfnamefont {P.}~\bibnamefont {Sciences}}, \bibinfo {author} {\bibfnamefont {S.~S.}\ \bibnamefont {Board}},\ and\ \bibinfo {author} {\bibfnamefont {C.}~\bibnamefont {on~Astrobiology Science Strategy for the Search for Life in~the Universe}},\ }\href@noop {} {\emph {\bibinfo {title} {An Astrobiology Strategy for the Search for Life in the Universe}}}\ (\bibinfo  {publisher} {National Academies Press},\ \bibinfo {year} {2019})\BibitemShut {NoStop}%
\bibitem [{\citenamefont {Jia}\ \emph {et~al.}(2018)\citenamefont {Jia}, \citenamefont {Kivelson}, \citenamefont {Khurana},\ and\ \citenamefont {Kurth}}]{jia2018evidence}%
  \BibitemOpen
  \bibfield  {author} {\bibinfo {author} {\bibfnamefont {X.}~\bibnamefont {Jia}}, \bibinfo {author} {\bibfnamefont {M.~G.}\ \bibnamefont {Kivelson}}, \bibinfo {author} {\bibfnamefont {K.~K.}\ \bibnamefont {Khurana}},\ and\ \bibinfo {author} {\bibfnamefont {W.~S.}\ \bibnamefont {Kurth}},\ }\bibfield  {title} {\bibinfo {title} {Evidence of a plume on europa from galileo magnetic and plasma wave signatures},\ }\href@noop {} {\bibfield  {journal} {\bibinfo  {journal} {Nature Astronomy}\ }\textbf {\bibinfo {volume} {2}},\ \bibinfo {pages} {459} (\bibinfo {year} {2018})}\BibitemShut {NoStop}%
\bibitem [{\citenamefont {Liao}\ \emph {et~al.}(2020)\citenamefont {Liao}, \citenamefont {Nimmo},\ and\ \citenamefont {Neufeld}}]{liao2020heat}%
  \BibitemOpen
  \bibfield  {author} {\bibinfo {author} {\bibfnamefont {Y.}~\bibnamefont {Liao}}, \bibinfo {author} {\bibfnamefont {F.}~\bibnamefont {Nimmo}},\ and\ \bibinfo {author} {\bibfnamefont {J.~A.}\ \bibnamefont {Neufeld}},\ }\bibfield  {title} {\bibinfo {title} {Heat production and tidally driven fluid flow in the permeable core of enceladus},\ }\href@noop {} {\bibfield  {journal} {\bibinfo  {journal} {Journal of Geophysical Research: Planets}\ }\textbf {\bibinfo {volume} {125}},\ \bibinfo {pages} {e2019JE006209} (\bibinfo {year} {2020})}\BibitemShut {NoStop}%
\bibitem [{\citenamefont {Shock}\ and\ \citenamefont {Holland}(2007)}]{shock2007quantitative}%
  \BibitemOpen
  \bibfield  {author} {\bibinfo {author} {\bibfnamefont {E.~L.}\ \bibnamefont {Shock}}\ and\ \bibinfo {author} {\bibfnamefont {M.~E.}\ \bibnamefont {Holland}},\ }\bibfield  {title} {\bibinfo {title} {Quantitative habitability},\ }\href@noop {} {\bibfield  {journal} {\bibinfo  {journal} {Astrobiology}\ }\textbf {\bibinfo {volume} {7}},\ \bibinfo {pages} {839} (\bibinfo {year} {2007})}\BibitemShut {NoStop}%
\bibitem [{\citenamefont {Cockell}\ \emph {et~al.}(2024)\citenamefont {Cockell}, \citenamefont {Simons}, \citenamefont {Castillo-Rogez}, \citenamefont {Higgins}, \citenamefont {Kaltenegger}, \citenamefont {Keane}, \citenamefont {Leonard}, \citenamefont {Mitchell}, \citenamefont {Park}, \citenamefont {Perl} \emph {et~al.}}]{cockell2024sustained}%
  \BibitemOpen
  \bibfield  {author} {\bibinfo {author} {\bibfnamefont {C.~S.}\ \bibnamefont {Cockell}}, \bibinfo {author} {\bibfnamefont {M.}~\bibnamefont {Simons}}, \bibinfo {author} {\bibfnamefont {J.}~\bibnamefont {Castillo-Rogez}}, \bibinfo {author} {\bibfnamefont {P.~M.}\ \bibnamefont {Higgins}}, \bibinfo {author} {\bibfnamefont {L.}~\bibnamefont {Kaltenegger}}, \bibinfo {author} {\bibfnamefont {J.~T.}\ \bibnamefont {Keane}}, \bibinfo {author} {\bibfnamefont {E.~J.}\ \bibnamefont {Leonard}}, \bibinfo {author} {\bibfnamefont {K.~L.}\ \bibnamefont {Mitchell}}, \bibinfo {author} {\bibfnamefont {R.~S.}\ \bibnamefont {Park}}, \bibinfo {author} {\bibfnamefont {S.~M.}\ \bibnamefont {Perl}}, \emph {et~al.},\ }\bibfield  {title} {\bibinfo {title} {Sustained and comparative habitability beyond earth},\ }\href@noop {} {\bibfield  {journal} {\bibinfo  {journal} {Nature Astronomy}\ }\textbf {\bibinfo {volume} {8}},\ \bibinfo {pages} {30} (\bibinfo {year} {2024})}\BibitemShut {NoStop}%
\bibitem [{\citenamefont {Pradhan}\ and\ \citenamefont {Pokhrel}(2013)}]{pradhan2013spectrophotometric}%
  \BibitemOpen
  \bibfield  {author} {\bibinfo {author} {\bibfnamefont {S.}~\bibnamefont {Pradhan}}\ and\ \bibinfo {author} {\bibfnamefont {M.~R.}\ \bibnamefont {Pokhrel}},\ }\bibfield  {title} {\bibinfo {title} {Spectrophotometric determination of phosphate in sugarcane juice, fertilizer, detergent and water samples by molybdenum blue method},\ }\href@noop {} {\bibfield  {journal} {\bibinfo  {journal} {Scientific World}\ }\textbf {\bibinfo {volume} {11}},\ \bibinfo {pages} {58} (\bibinfo {year} {2013})}\BibitemShut {NoStop}%
\end{thebibliography}%

\newpage
\clearpage

\subsection*{Methods}

A complete list of the materials used for this work is provided in the Supplementary Information.\\

\noindent \textbf{Preparation of spherical microspheres}\\
To prepare the microspheres, a standard protocol was followed as described: in a 50 mL centrifuge tube, 2 mL of 100 mM diammonium molybdate solution was added, followed by adding 5 mL of 1000 mM diammonium hydrogen phosphate, and 1 mL of 100 mM iron sulfate. The mixture was shaken vigorously, followed by adding hydrochloric acid at a final concentration of 450 mM. The solution was mixed once again. Lastly, the sole carbon source, formaldehyde was added at a final concentration of 3 M, followed by mixing it thoroughly. The total volume of the mixture was made up to 20 mL by adding water. The prepared mixture was then either partitioned into microcentrifuge tubes or deep well plates and incubated under ambient well-lit conditions at 25\degree C or in the dark where relevant. All experiments for the microsphere analysis were done using replicates of these samples at specific time intervals. The phase space was systematically spanned by varying the starting concentrations of reagents in deep-well plates, which were imaged using a flat-bed scanner at specific time intervals. 50 $\mu$L of sample was pipetted out from each well for microscopy imaging.\\

\noindent \textbf{Preparation of wells for imaging}\\
Chambers made from poly(dimethyl siloxane) (PDMS) were prepared to enable imaging of microspheres at smaller volumes. These chambers allowed for precise control over the density of the microspheres and minimized rapid evaporation of the solution during long-term imaging experiments. To prepare the chambers, a plastic container was used, and PDMS base was mixed with its crosslinker in a 10:1 ratio by weight. The mixture was stirred vigorously until bubbles formed. The resulting mixture was poured over masks placed in 30 mm Petri dishes. To remove trapped air bubbles, the setup was degassed in a desiccator for 1-2 hours. Once degassed, the PDMS was cured in an oven at 70\degree C overnight. The cured PDMS sheets were cut into cuboids using a scalpel and cleaned thoroughly with isopropyl alcohol. Holes with a 2 mm radius were punched into these cuboids and were plasma bonded to confocal dishes (Biofil 15 mm, non-treated dishes) to ensure a strong seal thereby creating a ``well'' for reactions. To finalize bonding, the plasma treatment was continued for 15 minutes. The coverslips and PDMS blocks were then removed, and the exposed sides were quickly aligned and pressed together to bond them securely. These bonded chambers were stored at ambient temperature until use. For imaging, the microspheres solution was loaded into the prepared PDMS wells, and analysed under the microscope.\\

\noindent \textbf{Protocol for microscopy imaging of the spherical microspheres}\\
For microscopy imaging, 50 $\mu$L of the mixture was transferred into confocal dishes equipped with a PDMS well, and incubated at 25\degree C under ambient lighting conditions. Independent samples were imaged at specific time points by loading onto a microscope. Image acquisition was carried out using an Olympus FV3000 Confocal Microscope with a 60x oil objective. This setup enabled the detailed visualization of the microspheres. The color images were taken with an Olympus FV1000 Microscope, equipped with an RGB camera.\\

\noindent \textbf{Energy Dispersive X-Ray Spectroscopy (EDX)}\\
For Energy Dispersive X-Ray Spectroscopy (EDX) analysis, aqueous mixtures forming the microspheres were prepared as described previously. After 24 hours of incubation, the samples were transferred onto a round coverslip and allowed to stick. This was followed by desiccation for 24 hours to eliminate moisture. The surface of the dried sample was then cleaned using a plasma cleaner to ensure the removal of contaminants. For time-dependent EDX measurements, multiple replicates were incubated and dried at specific time intervals \textit{i.e.} 1 hour, 2 hours, 4 hours and 8 hours after reaction initiation, followed by plasma cleaning.

For EDX measurements, a selected region of interest (ROI) of the size of microspheres were scanned, analyzing the emitted X-ray spectra to determine the elemental composition. This allowed the identification of key elements present on the surface of the microspheres, such as oxygen, carbon, nitrogen, and metal ions, providing insights into their chemical makeup. The resulting data, normalized by area of the ROIs, highlighted the proportions of each element and any deviations from the starting concentrations of the reagents used in the reaction mixture via selective enrichment.\\

\begin{table}[h!] 
\centering 
\caption{Elemental composition by normalized weight percentages (\%). The data shows the control (initial solution) compared against five experimental replicates (microspheres).}
\label{tab:elemental_data} 
\begin{tabular}{lrrrrrr}
\toprule
& & \multicolumn{5}{c}{Experimental Replicates (Microspheres)} \\
\cmidrule(l){3-7} 
Element & \multicolumn{1}{c}{Control} & \multicolumn{1}{c}{Rep 1} & \multicolumn{1}{c}{Rep 2} & \multicolumn{1}{c}{Rep 3} & \multicolumn{1}{c}{Rep 4} & \multicolumn{1}{c}{Rep 5} \\
\midrule
Carbon     & 21.23 & 14.07 & 13.96 & 12.45 & 14.32 & 15.87 \\
Nitrogen   & 4.14  & 12.11 & 11.63 & 9.89  & 10.74 & 14.81 \\
Oxygen     & 39.26 & 30.95 & 29.00 & 24.49 & 28.32 & 35.47 \\
Phosphorus & 5.52  & 3.70  & 3.14  & 2.78  & 2.77  & 1.95  \\
Molybdenum & 1.16  & 37.90 & 40.68 & 49.19 & 42.38 & 30.79 \\
Iron       & 0.35  & 0.59  & 1.37  & 0.48  & 0.49  & 0.75  \\
Chlorine   & 28.34 & 0.69  & 0.22  & 0.72  & 0.98  & 0.35  \\
\bottomrule
\end{tabular}
\end{table}

\noindent \textbf{Calculation of the starting elemental composition of the aqueous mixture}\\

The starting elemental composition of the aqueous mixture was calculated from known starting concentrations. Since, total volume is constant for each chemical compound, the number of moles of each compound is directly proportional to their molarities. Given the atomic weight of each element, the normalized weight percentage is as follows-

The mixture under consideration has five solutes: ammonium molybdate (10~mM), iron(II) sulfate (5~mM), diammonium hydrogen phosphate (250~mM), hydrochloric acid (450~mM), and formaldehyde (3~M). The molecular formulas used for the calculations were based on the most commonly encountered hydrated or anhydrous forms: \((\mathrm{NH}_4)_6\mathrm{Mo}_7\mathrm{O}_{24}\) for ammonium molybdate, \(\mathrm{FeSO}_4\) for iron sulfate, \((\mathrm{NH}_4)_2\mathrm{HPO}_4\) for diammonium phosphate, \(\mathrm{HCl}\) for hydrochloric acid, and \(\mathrm{CH}_2\mathrm{O}\) for formaldehyde. For each compound, the elemental composition (i.e., number of atoms of each element of interest) was determined from the chemical formula. For each compound, the mass of each relevant element contributed per liter of solution was calculated using the equation:\[\text{mass}_{\text{element}} = C_{\text{compound}} \times n_{\text{element}} \times M_{\text{element}}\]where \( C_{\text{compound}} \) is the molar concentration of the compound in mol/L, \( n_{\text{element}} \) is the number of atoms of the element per molecule of the compound, and \( M_{\text{element}} \) is the atomic mass of the element (in g/mol). This calculation yields the absolute mass of each element per liter of solution. The contributions of all compounds were summed for each element to obtain the total mass of each element per liter. These values were then normalized for the seven target elements so that their total mass was added to 100\%. This was done by dividing each elemental mass by the sum of all elemental masses and multiplying by 100 to obtain the weight percentage as follows:\[\text{Normalized weight \%} = \left( \frac{\text{mass}_{\text{element}}}{\sum \text{mass}_{\text{all target elements}}} \right) \times 100\]

\begin{table}[h!]
\centering
\caption{Normalized weight percentages of selected elements in the aqueous mixture.}
\begin{tabular}{l c}
\hline
\textbf{Element} & \textbf{Normalized Weight \%} \\
\hline
C  & 25.41 \\
N  & 5.53 \\
O  & 53.99 \\
P  & 0.55 \\
Mo & 4.06 \\
Fe & 0.20 \\
Cl & 11.25 \\
\hline
\textbf{Total} & \textbf{100.00} \\
\hline
\end{tabular}
\label{tab:element_weights}
\end{table}

\noindent \textbf{Scanning Electron Microscopy (SEM)}\\
The microspheres were prepared as described previously. The sample slides for Scanning Electron Microscopy (SEM) imaging were prepared the same way as described in the Energy Dispersive X-Ray Spectroscopy (EDX) protocol. All the samples were incubated for 24 hours before drying. Imaging was performed using a Carl Zeiss Ultra 55 Field Emission Scanning Electron Microscope (FESEM).\\

\noindent \textbf{Transmission Electron Microscopy (TEM)}\\
For Transmission Electron Microscopy (TEM) imaging the 24 hours samples were drop-cast on grids, followed by sonication for 30-45 minutes, and then kept for drying under IR lamp for 4-5 hours and then desiccated overnight. Imaging was performed using a Titan Themis 300kV from FEI (now Thermo) Transmission Electron Microscope. The imaging parameters (accelerating voltage, working distance, magnification) were optimized to capture high-resolution images of the structural details of the microspheres.

For the TEM cross-section protocol, the sample was mixed with 50 \% ethanol for 15 min, 70 \% ethanol for 15 min, 90 \% ethanol for 15 min, followed by 100 \% ethanol three times for 15 min each. The sample was then mixed with 100 \% ethanol \& Epon 812 resin (1:1) and the solvent was evaporated in a rotator over night. The sample was then mixed with pure resin and kept for 6 hours, followed by embedding the sample into moulds. It was then polymerised at 60\degree C for 48 hours. The sample were cut into fine cross-sections of 70 nm, loaded onto a grid and imaged under the microscope.\\

\noindent \textbf{Oxidation of the microspheres for observing internal structuration}\\
The microspheres solution was prepared as described in the previous sections. The particles were pipetted out post 24 hours incubation and treated with an aqueous mixture of 2 \% hydrogen peroxide solution and 3 M sodium hydroxide solution for controlled oxidation of the microspheres. These were then imaged using an Olympus FV1000 Microscope, equipped with an RGB camera.\\

\noindent \textbf{Bursting of the microspheres on optical stimulation}\\
The standard protocol was followed to prepare the microspheres. 50 $\mu$L of the mixture was immediately transferred to an imaging dish equipped with a PDMS chamber; it was then sealed to reduce evaporation. The chambers were incubated for 8 hours at 25\degree C under ambient lighting conditions. The samples were then imaged using an Olympus FV3000 Confocal Microscope with a 60x oil objective lens. For the stimulation imaging protocol, the power of the LASER (405 nm) was increased in a step-wise manner upto 70 \% of the maximum intensity. In parallel, time lapses were captured in the Differential Interference Contrast (DIC) mode to observe the effect of increasing coherent light stimulation on the microspheres.\\

\noindent \textbf{Atomic Force Microscopy (AFM)}\\
The starting aqueous mixture was prepared as described in the previous sections. The samples were incubated for either 4 hours or 24 hours. Post incubation, the microsphere samples were diluted to 1:10 or 1:20 in filtered salt buffer. A 20 $\mu$L drop of the diluted sample was placed onto fresh silica substrate and left to incubate at room temperature for 10-20 minutes. To remove any detached particles, 100 $\mu$L of filtered salt buffer was applied to the slide and removed gently using a fresh tissue. The washing step was repeated 2-4 times. The sample was then dried at room temperature for 5-10 minutes. The prepared slides were then analysed using the Oxford instruments- Asylum Research Model MFP3D Bio. Measurements were performed in the pointscan tapping mode with an ACSTA tip, using a scan rate of 0.5 Hz with a velocity of 1 $\mu$m/sec.\\



\noindent \textbf{Spectrophotometry}\\
Absorbance measurements were conducted to monitor changes in the optical density (OD) during the formation of the microspheres. Freshly prepared reaction mixture, along with control (without formaldehyde), was pipetted into 24-well plates in replicates. Absorbance scan was performed using a Tecan SPARK Spectrophotometer, set to measure from 200 nm to 1000 nm at a step size of 5 nm. Measurements were carried out using a kinetic loop at 25\degree C with constant shaking over time. This setup provided the absorption spectrum, including the peak wavelength and any spectral shifts over time and wavelength.\\

\noindent \textbf{Growth of individual microspheres}\\
For time-lapse imaging, growth movies were captured to monitor the dynamics of single microspheres over time. The reaction mixture was prepared as mentioned above, and 50 $\mu$L of the sample was immediately transferred to the imaging dishes integrated with PDMS wells, followed by sealing the chamber to reduce evaporation. Imaging was initiated with a time-series protocol on the Olympus FV3000 Confocal Microscope with a 60x oil objective lens in the DIC mode. Frames were captured at regular intervals over 8 hours. The small volume of the PDMS wells allowed precise observation of individual microspheres' growth dynamics during this period. The time lapses were further analysed via ImageJ and in-house MATLAB codes for quantification.\\

\noindent \textbf{Mass measurements}\\
Multiple replicates of independent reaction and control mixtures were prepared as described previously. The control mixtures were prepared without formaldehyde, keeping the concentration of all the other ingredients unchanged. Whole centrifuge tubes were taken at specific time points and analysed. Each sample was filtered out to strain the liquid and the filter papers were dried in a desiccator overnight. After drying the filter papers were weighed. The dry mass of the samples were estimated by subtracting the weight of the filter papers before and after filtration (and drying).\\

\noindent \textbf{Capacity for self-perpetuation of the microspheres}\\
The reaction mixture was prepared as described in the preceding section. Immediately after preparation, 50 $\mu$L of the sample was transferred into imaging dishes fitted with PDMS wells, ensuring consistent sample height and confinement. The chambers were then sealed to minimize evaporation throughout the experiment. Time-lapse imaging was performed for 8 hours using the same imaging parameters and acquisition protocol described earlier. After this initial imaging period, the samples were subjected to the laser-bursting (optical stimulation) protocol, during which the 405 nm laser was operated at 90 \% power to induce complete bursting of the microspheres. Following optical stimulation, the samples were returned to the microscope and imaged for an additional 8 hours under identical acquisition conditions to monitor any regrowth, reformation, or self-propagation of the released microspheres. At the end of the imaging sequence, the chamber assembly was carefully dismantled. Coverslips containing the residual material were prepared for EDX analysis using the procedure described previously. For comparison, EDX spectra of intact microspheres were also recorded from regions within the same sample but spatially well-separated from the laser-bursting zone. This allowed direct assessment of changes in elemental composition associated specifically with the bursting and regrowth cycle. In total, five replicates were performed for each condition.\\

\noindent \textbf{pH measurements}\\
To investigate the chemical activity of the microsphere mixtures, the bulk pH of the samples were measured over time. The measurements were carried out for both microsphere mixtures and controls. The controls were prepared by omitting one of the key ingredients at a time, such as ammonium molybdate, diammonium hydrogen phosphate, iron sulfate, or formaldehyde, keeping the total concentration of all the other ingredients same. Additionally, a control without both ammonium molybdate and iron sulfate was used to observe the pH changes in the absence of these metal catalysts. Another control was prepared to study the effect of hydrochloric acid on formaldehyde in the absence of other reagents. Using a standard pH meter, pH readings were taken as a function of time. The measured pH values were converted to [H+] ion concentration using the equation: [H+] = 10\textsuperscript{-pH}.\\

\noindent \textbf{Calorimetry}\\
To measure the amount of heat flow in the system, both microspheres and control (without formaldehyde) reaction mixtures were prepared similarly as described previously. From the freshly prepared solutions, 2 mL was immediately transferred to ampules specifically designed for calorimetry measurements. The ampules were sealed and transferred to the TAM IV Calorimeter (TA Instruments). For all samples and controls, isothermal calorimetry measurements were then carried out at the same incubation temperature. The heat flow as a function of time was integrated to quantify the total heat generated over time.\\


\noindent \textbf{Mass Spectrometry (MS)}\\
To assess the evolving chemical complexity of the reaction mixtures and identify molecular species synthesized over time, mass spectrometry analyses were performed on both time-series ampoule samples and large-scale microsphere forming solutions. 

For the time-resolved dataset, reaction mixtures containing \textsuperscript{13}C-labelled formaldehyde and appropriate controls lacking formaldehyde were prepared in 1 mL volumes inside sterile glass ampoules. The ampoules were sealed with caps and incubated at 25 \degree C in a class 10000 clean room, under controlled light conditions using an LCS-100 Model Solar Simulator. The LCS-100 Model was equipped with an AM1.5G filter which generated 1 Sun irradiance. The working distance was adjusted to allow continued 1 Sun operation over the time course of the experiment. Ampoules were collected at predefined time points and immediately stored at $\text -20$\degree C until further analysis.

In parallel, to investigate the molecules formed in the microspheres and the supernatant solution separately, microsphere forming reaction mixtures were prepared on a larger 100 mL scale using the same chemical composition described previously. These mixtures were set up in sealed round-bottom flasks and incubated under identical solar simulator conditions and separately in a transparent sealed box under natural sunlight-night conditions. For the solar simulator samples, the formaldehyde consisted of unlabelled \textsuperscript{12}C formaldehyde (99\%), spiked with (1\%) \textsuperscript{13}C-labelled formaldehyde to facilitate isotope tracing. No \textsuperscript{13}C-labelled formaldehyde was used for the sunlight incubated samples. Whole sample aliquots were collected at the designated time points and stored at $\text -20$\degree C prior to downstream processing. 

Additionally, separate 100 mL batches of the reaction mixture were prepared in sealed transparent glass containers and placed inside an in-house-built irradiation box. These were exposed to natural sunlight for 60 days to mimic long-term environmental photochemistry. At the end of the exposure period, samples were collected and stored at $\text -20$\degree C until further processing. 

For the mass spectrometry sample preparation step, all frozen samples were thawed and 1 mL of each collected sample was transferred to microcentrifuge tubes and centrifuged at 11,000g to pellet and aggregate the microspheres. The resulting supernatants were transferred into fresh tubes. All samples, both supernatants and pellet-containing fractions, were then acidified with hydrochloric acid to dissolve and digest the microspheres and to maintain uniform chemical conditions across all samples prior to liquid chromatography injection. The small volume time series ampoule samples were processed directly in the acidification step without the centrifugation step, keeping the protocol consistent.

All the acidified controls, samples and supernatants were then subjected to reverse-phase column purification using Strata-X solid-phase C-18 column cartridges. The column was equilibrated by passing 1 mL of LC-MS grade methanol under vacuum, followed by washing with LC-MS grade water. The entire volume of the samples were passed through the columns, and the columns were washed with 2\% LC-MS grade methanol. Elution was carried out using a buffer consisting of 2\% formic acid in methanol, and the eluted material was collected in centrifuge tubes. The eluates were dried using a rotary evaporator under vacuum. Finally, the dried samples were injected into a coupled liquid chromatography mass spectrometry (LC-MS) Thermo LTQ Orbitrap XL equipment for analysis of their chemical composition. The column used for elution was a C18 Aquity column (2.1x100 mm, 1.7 $\mu$m CSH). Injection volumes were scaled upto five times for concentration dependent confirmation of molecules. All generated data were analysed using Xcalibur, MS-DIAL and OpenMS software.\\ 

\noindent \textbf{Mass Spectrometry (MS) Analysis}\\
We infused our reaction mixture with $\sim$99\% \textsuperscript{13}C-formaldehyde to trace the incorporation of this labeled carbon into newly synthesized molecules. For a molecule with $n$ carbons built from this tracer, the dominant isotopologue is fully \textsuperscript{13}C-labeled (mass M), while the $\sim$1\% \textsuperscript{12}C impurity produces a diagnostic M$-$1 peak from single \textsuperscript{12}C substitutions. This M$-$1 signal is impossible for natural-abundance molecules (no isotope lighter than \textsuperscript{12}C exists), providing an unambiguous signature of \textit{de novo} synthesis.

The expected M$-$1/M intensity ratio follows from binomial statistics:
\begin{equation}
\frac{I_{\text{M$-$1}}}{I_{\text{M}}} = \frac{n(1-p)}{p} \approx \frac{n}{99}
\end{equation}
where $p = 0.99$ is the tracer purity. This allows estimation of carbon count: $n \approx 99 \times (I_{\text{M$-$1}}/I_{\text{M}})$.

LC-MS data were processed using OpenMS for feature detection and alignment. For each candidate anchor peak (putative M), we searched for a co-eluting M$-$1 partner at $m/z_{\text{M}} - 1.003355/z$ within $\pm$10~ppm and $\pm$10~s retention time tolerance.

Candidates passed the following filters:
\begin{enumerate}
    \item \textit{Intensity threshold}: Blank-subtracted anchor intensity $>$5000 counts
    \item \textit{M$-$1/M ratio bounds}: $0.005 \leq I_{\text{M$-$1}}/I_{\text{M}} \leq 0.30$
    \item \textit{Carbon check}: Estimated $n \leq m_{\text{neutral}}/14$~Da (physical limit for organic molecules)
    \item \textit{M$-$2 consistency}: $I_{\text{M$-$2}} < I_{\text{M$-$1}}$ (binomial expectation)
    \item \textit{M$+$1 depletion}: $I_{\text{M$+$1}}/I_{\text{M}} \leq 0.10$
\end{enumerate}

The M$+$1 filter exploits the fact that for fully \textsuperscript{13}C-labeled molecules, M$+$1 arises only from minor isotopes of other elements (\textsuperscript{2}H, \textsuperscript{15}N, \textsuperscript{17}O), yielding M$+$1/M $\approx$1--3\%. Natural-abundance molecules of similar size would show M$+$1/M $\approx n \times 1.1\%$ from \textsuperscript{13}C. Candidates exceeding 10\% are rejected as false positives or partially labeled species.

Detection was performed independently at each time point (Days 2 and 21), with the \textit{emergence day} defined as the first time point passing all criteria.\\

\noindent \textbf{Detection of \textit{De Novo} Synthesized Molecules via \textsuperscript{13}C Isotope Analysis and MS2 patterning}\\
To track carbon flux from the C1 feedstock into newly formed products, we supplemented the reaction with \textsuperscript{13}C-labelled formaldehyde. For the large-volume solar-simulator experiments used for molecular identification, the formaldehyde pool consisted predominantly of unlabeled (\textsuperscript{12}C) formaldehyde spiked with $\sim$1\% \textsuperscript{13}C-formaldehyde, enabling isotope tracing while preserving the native bulk chemistry of the mixture. The reaction mixtures incubated under natural day-night cycling had only unlabelled formaldehyde. As mentioned in the previous MS experimental protocol section, reaction aliquots were collected at designated time points, and for large-volume runs the microsphere-enriched (pellet) fraction was analyzed separately from the supernatant. Wherever required, injection volumes were scaled (up to five-fold) for concentration-dependent confirmation.

MS data were processed using Xcalibur and MS-DIAL software for feature detection, alignment, and annotation. Putative identities were first obtained by matching MS1 features (m/z, retention time, adduct type) to reference databases, and then strengthened by MS2 spectral matching against reference fragmentation signatures. We applied stringent mass-matching tolerances of 0.01 Da for MS1 and 0.025 Da for MS2, ensuring that only close matches were retained for downstream interpretation. For the curated molecule lists reported in the Supplementary tables, a match score threshold ($>$0.90) was used to filter high-confidence candidates. Representative examples of the MS1 isotope patterns and the corresponding MS2 matches (experimental spectra compared against database reference spectra) are provided in SI Figs. \ref{fig:s4_lab} and \ref{fig:s4_sunlight}, while the full filtered candidate sets are summarized in Table \ref{tab_lab} (solar simulator; spiked with $\sim$1\% \textsuperscript{13}C-formaldehyde) and Table \ref{tab_daynight} (day–night cycling experiment). 

In the isotope-spiked datasets, for each candidate we applied tight retention time tolerances followed by a series of filters (blank-subtracted and control subtracted intensities, expected isotope-ratio bounds, carbon-count checks, and consistency of higher isotopologues) to reject false positives and carry out molecular species identification. We then leveraged the appearance of paired isotopologues, an unlabeled M (\textsuperscript{12}C) peak and the corresponding M+1 (\textsuperscript{13}C) peak, to support incorporation of labelled carbon into products detected in the microsphere-enriched phase (SI Fig. \ref{fig:s4_lab}). Together, the combined MS1 isotope evidence and MS2 fragmentation confirmation provide a conservative workflow for identifying candidate molecules whose presence is consistent with incorporation of carbon from formaldehyde into newly generated chemical products.\\


\noindent \textbf{Nuclear Magnetic Resonance (NMR)}\\
To assess the evolving chemical complexity of the reaction mixtures and to identify molecular species synthesized over time, Nuclear Magnetic Resonance (NMR) analyses were performed on the time-series ampoule samples containing $\sim$99\% \textsuperscript{13}C-formaldehyde as the sole C1 source.

As previously described in the MS sample preparation section, for the time-resolved dataset, reaction mixtures containing \textsuperscript{13}C-labelled formaldehyde were prepared in 1 mL volumes inside sterile glass ampoules. At each specific time points, the samples were purified and frozen at $\text -20$\degree C for further analysis. For the NMR measurements, 500 $\mu$l of each sample was mixed with 10 \% deuterated water and loaded into the NMR tubes. These were capped and loaded into a sample holder (spinner turbine) and were positioned correctly within the magnet's probe using a depth gauge. The instrument's pneumatic system was used to carefully lower the samples into the magnet, followed by shimming to adjust the magnetic field homogeneity. The measurements were made using a Bruker Avance-III HD 600 MHz instrument for which the RF (radio frequency) channel for \textsuperscript{13}C was used. Acquisition parameters were specifically adjusted for the measurements. For data analysis, Bruker's TopSpin software was used and the plots were generated using in-house Python codes.

\subsection*{Data availability}
All data that support the plots within this paper and other findings of this study are available from S.T. upon reasonable request and will be made available upon publication.

\subsection*{Code availability}
The computational methods that support the plots within this paper are described in the Supplementary Information and the underlying code is available from S.T. upon reasonable request and will be made available upon publication.

\subsection*{Acknowledgements}

We thank Anisha Shastri, Harshini Sangle, Malavika Anilkumar, Pathik Das, Sana Siroya and Tanzeel Ahmed for initial help with the experiments. We benefited from discussions and feedback from Jyotishman Dasgupta, Shachi Gosavi, Stephan Herminghaus, Madan Rao, Jack Szostak and Zorana Zeravcic. We thank the Central Imaging, Electron Microscopy (Angshuman Chowdhuri and Sunil Prabhakar), Nuclear Magnetic Resonance (Ranabir Das and Arnab Dey) and Mass Spectrometry (Nirpendra Singh and Ankit Jain) Facilities at NCBS, Bangalore. We thank Yatheendran K. M. from RRI, Bangalore for help with the Atomic Force Microscopy. We thank the Micro Nano Characterization Facility at CeNSE, IISc, Bangalore for help with the Scanning Electron Microscopy and Transmission Electron Microscopy. We acknowledge the Laboratory for Electro-Optical Systems (LEOS-ISRO) for experimental facility support. We acknowledge support from the Department of Atomic Energy (India), under project no.\,RTI4006, the Simons Foundation (Grant No.\,287975) and the Murty Trust.

\subsection*{Author contributions}

S.T. conceived the overall project. N.C. and S.T. designed the experiments. N.C. performed the experiments. N.C. and S.T. analysed the data and wrote the paper.

\subsection*{Competing interests}
The authors declare no competing interests.

\onecolumngrid
\setcounter{page}{0}
\makeatletter 
\renewcommand\thepage{S\arabic{page}}
\setcounter{figure}{0} 
\setcounter{equation}{0} 
\renewcommand{\thefigure}{S\@arabic\c@figure} 
\renewcommand{\thesection}{S\arabic{section}}
\renewcommand{\thesubsection}{S\arabic{subsection}}
\makeatother
\def\theequation{S\arabic{equation}}


\newpage

\section*{Supplementary Information}

\subsection{Materials}

All reagents used in this study were Analytical Reagent (AR) Grade. Ammonium Molybdate Tetrahydrate (CAS No. 12054-85-2), Iron Sulfate Heptahydrate (CAS No. 7782-63-0) and Formaldehyde solution (CAS No. 50-00-0) were purchased from Sigma-Aldrich. Diammonium Hydrogen Orthophosphate (CAS No. 7783-28-0) was bought from Thermo Fisher Scientific Inc. All the mineral acids used- Hydrochloric Acid (CAS No. 7647-01-0), Sulfuric Acid (CAS No. 7664-93-9) and Nitric Acid (CAS No. 7697-37-2) were also purchased from Thermo Fisher Scientific Inc. \textsuperscript{13}C-labelled Formaldehyde solution (CAS No: 3228-27-1) was purchased from Sigma-Aldrich. All experiments pertaining to the physical characterization of microspheres were carried out in Milli-Q water. All other chemical characterization and molecular identification experiments were carried out using Liquid Chromatography-Mass Spectrometry (LC-MS) grade water. LC-MS grade Water (CAS No. 7732-18-5) and LC-MS grade Methanol (CAS No. 67-56-1) were purchased from Thermo Fisher Scientific Inc. LC-MS grade Acetonitrile (CAS No. 75-05-8) was purchased from Biosolve Chimie. Reserpine Standard (CAS No. 50-55-5) for Liquid Chromatography-Mass Spectrometry (LC-MS) was bought from Sigma-Aldrich. The cartridge columns- Strata-X Polymeric Reversed Phase (Part No. 8B-S100-UBJ), used for Liquid Chromatography-Mass Spectrometry (LC-MS) sample preparation were purchased from Phenomenex. C18 Aquity column (2.1x100 mm, 1.7 $\mu$m CSH) was purchased from Waters India Pvt. Ltd.

\subsection{Supplementary Note: On the mechanism of microspherical compartments formation}

Here we present a plausible mechanism for the formation of the microspheres. Mo/Fe redox chemistry in an acidified phosphate environment generates Mo-blue–like polyoxometalate (POM) clusters and/or related Mo-oxide nanostructures, while formaldehyde-driven organic synthesis produces oligomeric and polymeric species in parallel. Together, these inorganic and organic components drive the emergence of hollow microspherical compartments whose boundary behaves as a partially arrested, viscoelastic shell/skin, a gelled interfacial network that selectively recruits Mo/N/Fe, resists fusion upon contact, hardens over time due to oxidation dependent cross-linking of the viscous gel material, and supports surface-catalyzed growth. As compartments mature, lumen chemistry produces smaller inclusions that are themselves growth-competent seeds, providing a minimal route to rudimentary perpetuation. 

This model satisfies the current experimental constraints. Firstly, selective elemental sequestration together with redox-linked blue/yellow polymorphs points to Mo-associated redox/cluster chemistry rather than a nonselective precipitation process (Fig. \ref{fig:fig1}; Fig. \ref{fig:fig5}\textbf{d}). The time-dependent spectrophotometry measurements (within the first two hours of starting the reaction mixture) reveal the emergence of a broad band around 840 nm, which is a signature for Mo-blue clusters doped with phosphate heteroatoms~\cite{pradhan2013spectrophotometric} (Supplementary Fig. \ref{fig:s3}\textbf{a}). Secondly, the observed hollow morphology, deformability, adhesion, and arrested fusion are consistent with a gel-like interface, rather than a purely rigid mineral precipitate (Fig. \ref{fig:fig2}\textbf{c, d}). Thirdly, the observation that growth slows with increasing radius and that polydispersity narrows over time supports a surface-catalyzed growth process, where growth is governed by the available reactive surface area (Fig. \ref{fig:fig3}\textbf{b} inset and associated discussion). Fourthly, the appearance of internal, growth-competent spherules provides direct evidence for a seeding-like pathway that can generate new compartments (Fig. \ref{fig:fig3}\textbf{h, i}). Finally, sustained calorimetric activity together with increasing molecular complexity indicates ongoing non-equilibrium chemistry that remains coupled to the compartments (Fig. \ref{fig:fig4}). 

While this framework offers a coherent interpretation of the current data, the precise chemical identities of the inorganic clusters and the initial organic products, the nature of the interfacial network, and the pathways linking lumen chemistry to seed formation remain open. These mechanistic aspects will be investigated in detail in future studies, including direct structural and redox-state characterization of Mo/Fe species, identification and localization of formaldehyde-derived oligomers/polymers, and perturbative experiments to dissect the processes behind growth, fusion arrest, and seeding.

\subsection{Supplementary Note: Hardening of compartments over time}

To quantify how the microsphere boundary evolves mechanically after formation, we performed time-dependent AFM measurements on particles harvested at early (4 hr) and later (24 hr) stages. In line with the force–distance responses reported in the main text, the compartments are soft, adhesive objects consistent with a deformable shell rather than a rigid precipitate (Fig. \ref{fig:fig2}\textbf{d}). When comparing the two time points, however, the AFM curves show a clear ``aging'' signature: microspheres at 24 hr are less compliant and exhibit a higher effective resistance to indentation than those at 4 hr, indicating progressive hardening of the shell with maturation (SI; Supplementary Fig. \ref{fig:s2}\textbf{d}).

One plausible interpretation is that the shell densifies and/or becomes increasingly crosslinked over time, e.g., through continued incorporation and reorganization of Mo-associated clusters together with formaldehyde-derived oligomeric/polymeric material, thereby shifting the boundary toward a more solid-like mechanical response while retaining strong adhesion. Functionally, such hardening would be expected to stabilize the hollow morphology, reinforce resistance to fusion/coalescence, and help maintain long-term compartment integrity while preserving deformability (Fig. \ref{fig:fig2}\textbf{c}).

Because particles were measured after adhesion to a substrate under identical handling for both time points, these AFM data are most robustly interpreted as a relative, time-dependent change in shell mechanics rather than an absolute modulus under native fully hydrated conditions. Nevertheless, the reproducible stiffening from 4 hr to 24 hr supports the conclusion that the shell is not a static precipitation product but an evolving, aging interface whose mechanical properties mature alongside the protocell’s chemical state.

\subsection{Supplementary Note: Formation of microspheres incubated in the dark}

Although several experiments in the main study were performed under ambient laboratory lighting or controlled solar-spectrum illumination (for reproducibility during long-term chemical characterization), in our chemical mixture, microsphere formation does not require light. We repeated the compartment formation and growth assays under dark conditions and confirmed that the compartments still form and grow, with kinetics comparable to those observed under ambient laboratory lighting (Supplementary Fig. \ref{fig:s3}\textbf{b}). Importantly, key signatures of ongoing non-equilibrium chemistry persist in the dark as well: the calorimetry measurements were performed under dark conditions, and as opposed to the control the reaction continues to show sustained energetic output and associated bulk chemical changes (Fig \ref{fig:fig4}\textbf{a}). Consistent with this, experiments performed in both ambient light and in the dark show that compartment growth is accompanied by proton production in the reaction mixture (Supplementary Fig. \ref{fig:s4}\textbf{a}).

Dark controls also support the conclusion that the chemical identity of the compartments is preserved in the absence of illumination. In Supplementary Fig. \ref{fig:s1}\textbf{f}, we present EDX measurements for microspheres incubated entirely in the dark for 24 hours (with corresponding SEM inset), demonstrating that robust microsphere formation under dark conditions still yields particles with the characteristic elemental signature of the system.

Together, these results show that, in our current experimental regime, light is not a necessary trigger for compartment formation. Instead, it likely acts primarily as a controllable external handle for standardization (and potentially modulation) of compartmental dynamics. Further, light-driven catalyse may drive chemical complexification along different paths eventually resulting in a different repertoire of synthesized molecules. Future work will systematically quantify how illumination conditions shift (or preserve) time-dependent composition, redox state, and material properties, separating what is intrinsically ``dark-robust'' from what is photo-modulated.

\subsection{Supplementary Note: Robust microspheres formation in the presence of additional components}

To probe whether microsphere formation is specific to a narrowly tuned ``clean'' recipe, or whether it persists robustly under more geochemically realistic ionic backgrounds, we repeated the protocell-forming reaction while supplementing the mixture with additional inorganic components (sodium, magnesium, potassium, and calcium salts). This robustness is particularly relevant in the context of the oceanic ``blue vacuoles'' comparison in Fig. \ref{fig:fig5}\textbf{a, b}. In the natural system, hollow, shell-like molybdenum-rich microspheres occur in an environment that is inherently salt-rich and compositionally complex, yet retain the characteristic morphology and elemental signature. 

As shown in Supplementary Fig. \ref{fig:s5}\textbf{b}, microspheres indeed form under these added-salt conditions, enabling direct SEM/EDX characterization at 24 hours post reaction initiation. The ability to recover well-defined compartments in the presence of these extra ions indicates that compartment formation is not a fragile artifact of a minimal laboratory mixture, but instead tolerates substantial ionic complexity. The naturally occurring oceanic structures and the synthetic compartments from our experiments share a molybdenum-dominated EDX signature with substantial carbon, oxygen, and phosphorus, and the time-evolving composition of synthetic protocells converges toward that of the oceanic microspheres (Fig. \ref{fig:fig5}\textbf{c}). Taken together, Supplementary Fig. \ref{fig:s5}\textbf{b} supports the idea that the protocell forming process can operate under more environmentally realistic conditions, consistent with the broader claim that similar chemistries could plausibly play out in natural salt-containing settings. 

\newpage
\clearpage


\begin{table}
\centering
\caption{Molecules present in the large volume solar simulator incubated sample. Match score of greater than 0.90 was selected as threshold.\label{tab_lab}}
\footnotesize
\begin{tabularx}{\textwidth}{rrrll}
\toprule
Database ID no & M/Z & RT (min) & Ion Type & Metabolite Name \\
\midrule
1511 & 235.11497 & 8.402 & [M+H]+ & Triethylene glycol diacetate \\
2289 & 246.24004 & 11.066 & [M+H]+ & SPB 14:0;2O \\
1512 & 252.14148 & 8.402 & [M+H]+ & Prosulfocarb \\
1447 & 286.1803 & 7.786 & [M+NH4]+ & NCGC00381044-01!3 \\
2558 & 290.26553 & 12.918 & [M+H]+ & SPB 16:0;3O \\
2705 & 309.12714 & 14.986 & [M+H]+ & Toddalolactone \\
1697 & 401.25763 & 9.625 & [M+H]+ & Bonactin \\
1520 & 402.22955 & 8.486 & [M+H]+ & NCGC00347551-02 \\
1540 & 446.2554 & 8.722 & [M+H]+ & NCGC00380693-01 \\
1806 & 489.31055 & 10.017 & [M+H]+ & SM 14:3;2O/5:0 \\
1832 & 494.29605 & 10.079 & [M-H2O+H]+ & Cysteine conjugated cholic acid \\
1917 & 496.2991 & 10.292 & [M+H]+ & PE 8:0\_10:0 \\
2101 & 530.28888 & 10.724 & [M+NH4]+ & Dibenzyl–diaminoindolyl triazonanetrione \\
1595 & 534.3067 & 9.125 & [M+NH4]+ & Tetramethyl dihydroxy tetracycle dimethylbutenoate \\
1925 & 560.33453 & 10.303 & [M+H]+ & PC 18:0 \\
1928 & 577.36169 & 10.303 & [M+NH4]+ & SL 14:3;O/18:5 \\
\bottomrule
\end{tabularx}
\end{table}

\clearpage
\newpage

\begin{table}
\centering
\caption{Molecules present in the large volume day-night cycle incubated sample. Match score of greater than 0.90 was selected as threshold.\label{tab_daynight}}
\begin{tabularx}{\textwidth}{rrrll}
\toprule
Database ID no & M/Z & RT (min) & Ion Type & Metabolite Name \\
\midrule
6278 & 83.06 & 25.08 & [M+H]+ & 1-Methylimidazole \\
581 & 102.13 & 1.38 & [M+H]+ & Hexylamine \\
2921 & 135.08 & 15.03 & [M+H-NH3]+ & 2-(4-Methoxyphenyl)ethylamine \\
2452 & 139.07 & 12.71 & [M+H]+ & 1,4-Dimethoxybenzene \\
753 & 148.06 & 3.21 & [M+H]+ & Glutamic acid \\
2477 & 152.02 & 11.63 & [M+H]+ & Thieno[3,2-b]pyridin-7-ol \\
1964 & 163.13 & 10.08 & [M+H]+ & Diethylene glycol monobutyl ether \\
6094 & 179.07 & 24.35 & [M+H]+ & 3-Methoxycinnamic acid \\
2661 & 180.10 & 13.88 & [M+H]+ & Fusaric acid \\
3343 & 194.12 & 16.34 & [M+H]+ & Ethyl 4-(dimethylamino)benzoate \\
2597 & 242.28 & 13.41 & [M+H]+ & Dioctylamine \\
6135 & 256.26 & 24.47 & [M+H]+ & 4-Dodecylmorpholine \\
4675 & 259.19 & 19.78 & [M+H]+ & Diethyl decanedioate \\
3423 & 358.37 & 17.00 & [M+H]+ & SPB 22:0;2O \\
3430 & 457.31 & 17.00 & [M+H]+ & Oleanoic Acid \\
\bottomrule
\end{tabularx}
\end{table}

\clearpage
\newpage

\begin{figure*}
\centering
\includegraphics[width=0.8\textwidth]{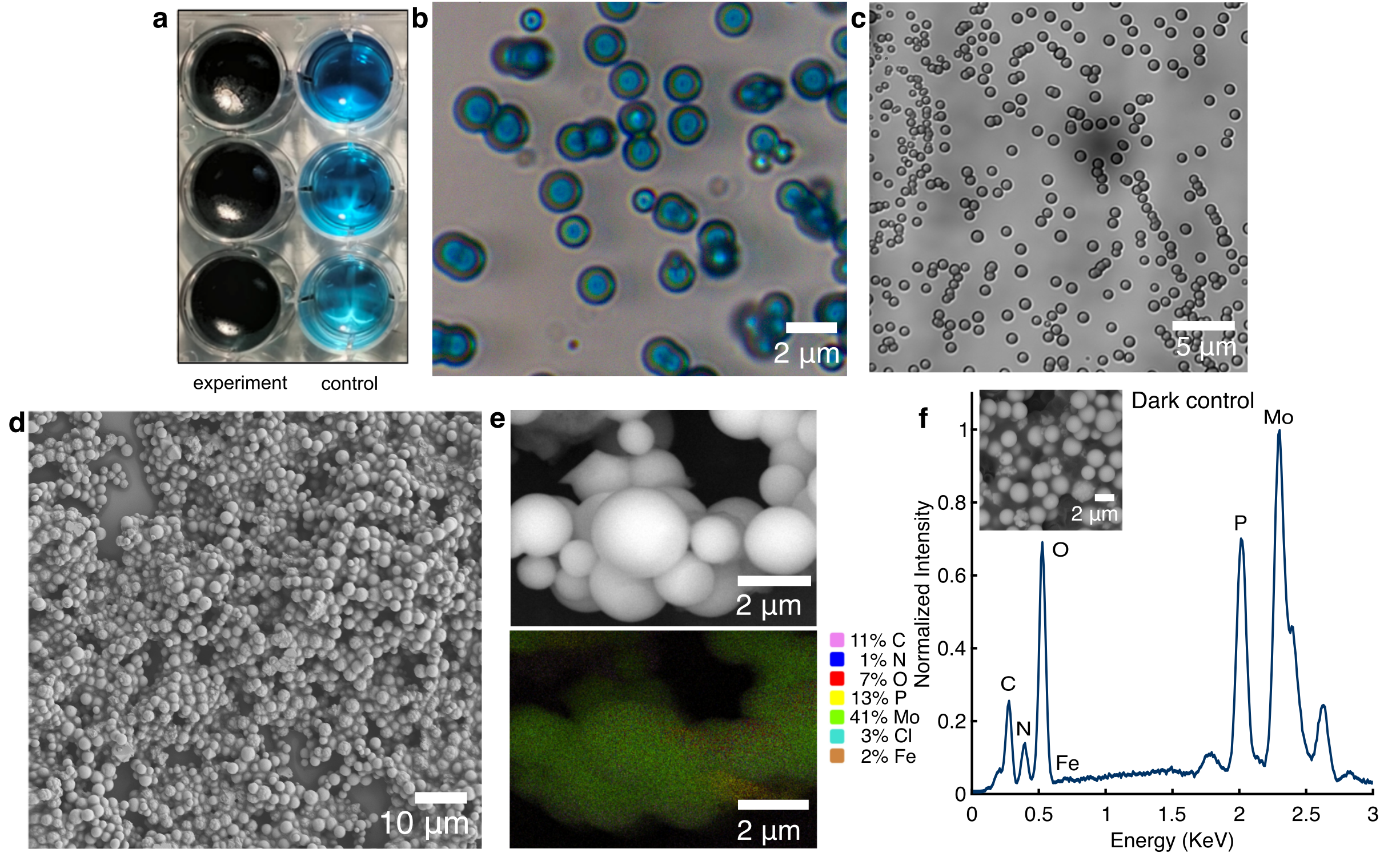}
\caption{\textbf{a}. Scanned images of the reactions carried out in deep well plates, imaged at 8 hours post incubation. The experiment reactions turned dark blue whereas the control reaction mixtures maintained a light blue color.
\textbf{b}. RGB image of the blue spheres at 24 hours post reaction initialisation.
\textbf{c}. Bright field images of the spherical microparticles revealing a monodisperse particle size distribution, 4 hours post reaction initialisation.
\textbf{d}. Scanning Electron Microscopy (SEM) imaging of the spherical microspheres, 24 hours after start of reaction.
\textbf{e}. SEM image (top panel) and corresponding EDX data represented as a pixel map (bottom panel) of the elemental composition of the spherical microparticles at 24 hours. Different colors highlight the spatial distribution of different elements.
\textbf{f}. EDX data of the microspheres incubated under dark conditions, 24 hours post initialisation. Inset shows SEM image for the same.
}
\label{fig:s1}
\end{figure*}

\begin{figure*}
\centering
\includegraphics[width=0.8\textwidth]{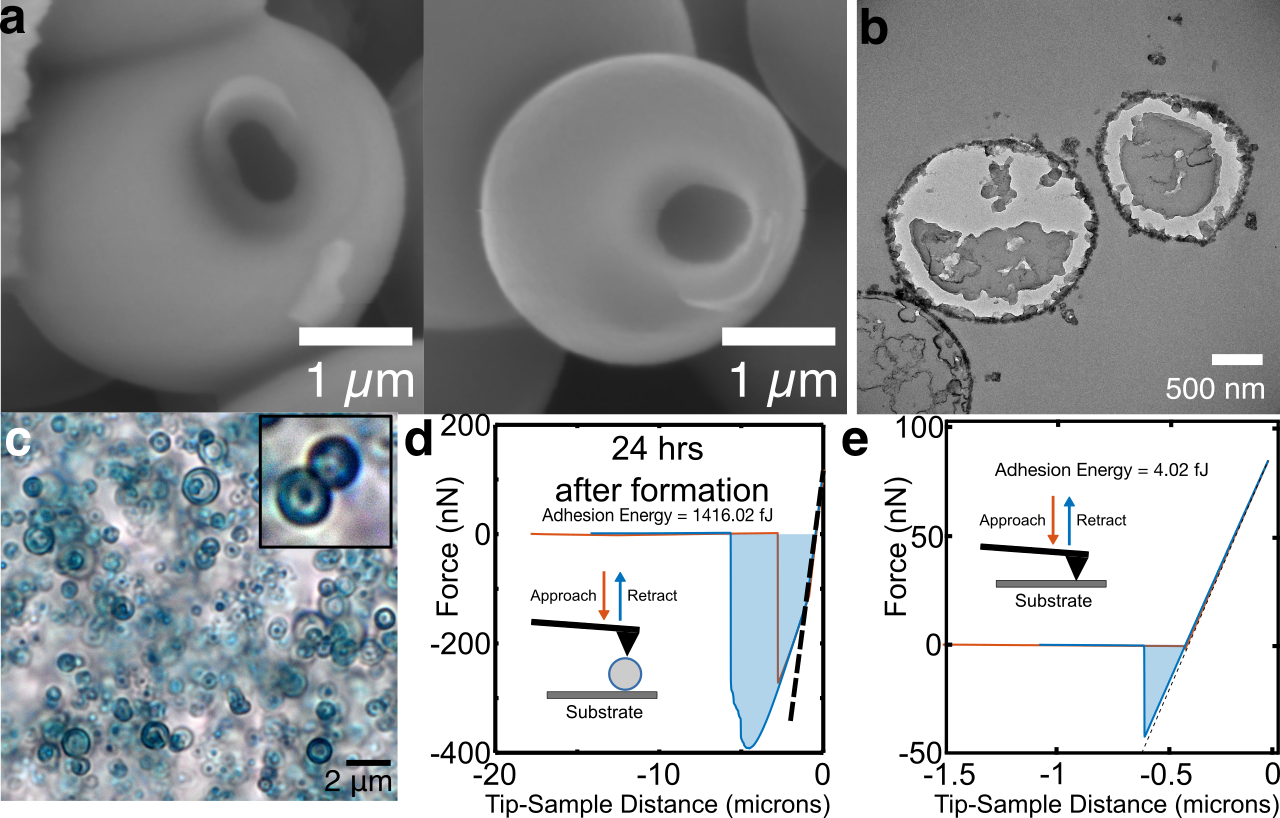}
\caption{\textbf{a}. SEM images of the microspheres at 24 hours reveal an internal hollow structure.
\textbf{b}. TEM cross-section image of the microspheres (24 hours post start of reaction) showcasing a boundary and heterogeneous lumen, highlighting their compartmental nature.
\textbf{c}. RGB color imaging of the blue microspheres post treatment with a solution of hydrogen peroxide and sodium hydroxide, thus partially oxidizing the molybdenum and iron in the particles. As a result, the internal structuration of the compartments can be observed.
AFM measurements quantification for the \textbf{d}. compartments at 24 hours post formation and \textbf{e}. glass coverslip (control surface).
}
\label{fig:s2}
\end{figure*}

\begin{figure*}
\centering
\includegraphics[width=0.8\textwidth]{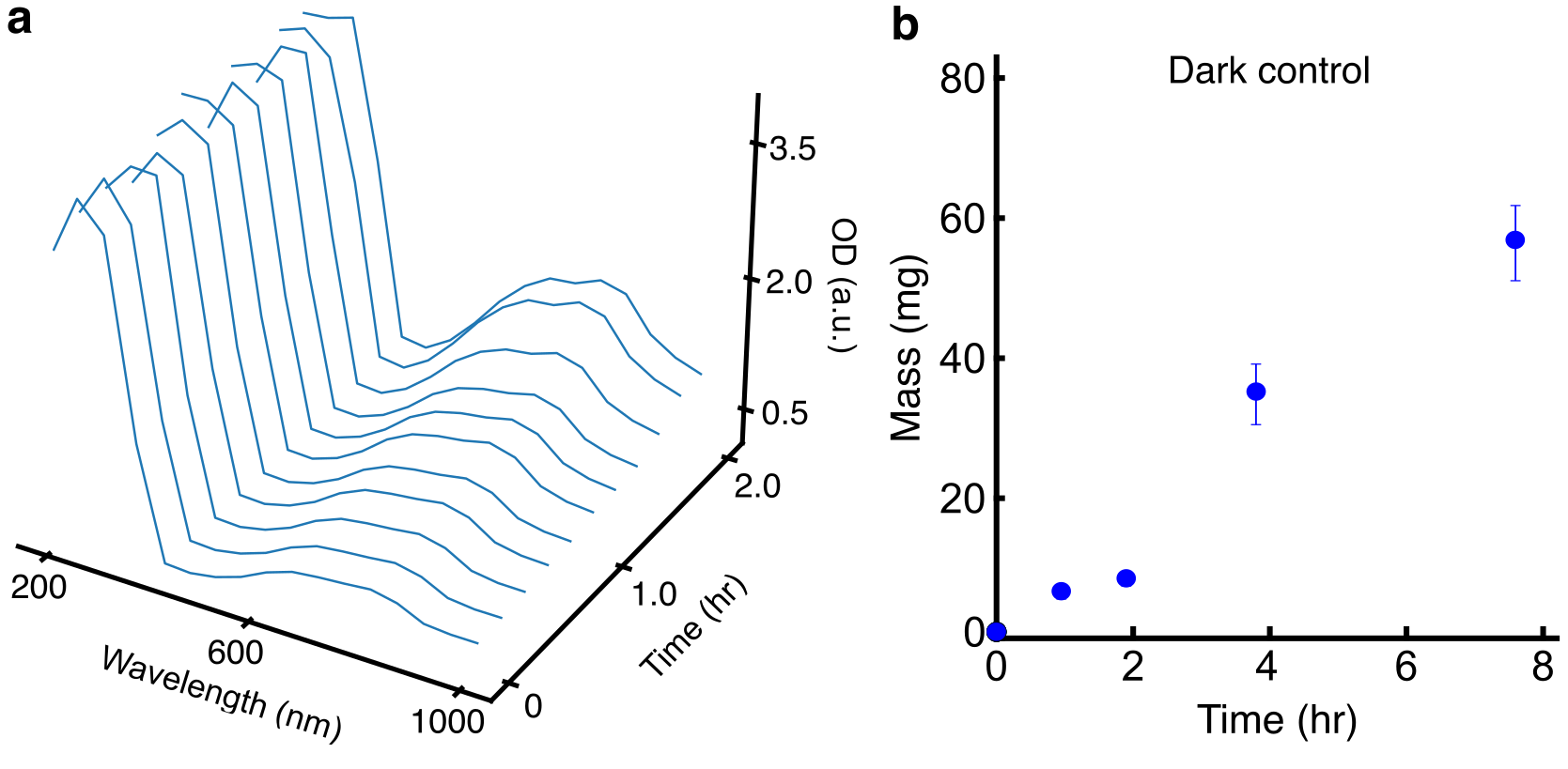}
\caption{\textbf{a}. Spectrophotometry measurements of the reaction mixture across wavelengths as a function of time.
\textbf{b}. Dry mass measurements of the microspheres incubated under dark conditions as a function of time. Error bars represent triplicates.
}
\label{fig:s3}
\end{figure*}

\begin{figure*}
\centering
\includegraphics[width=\textwidth]{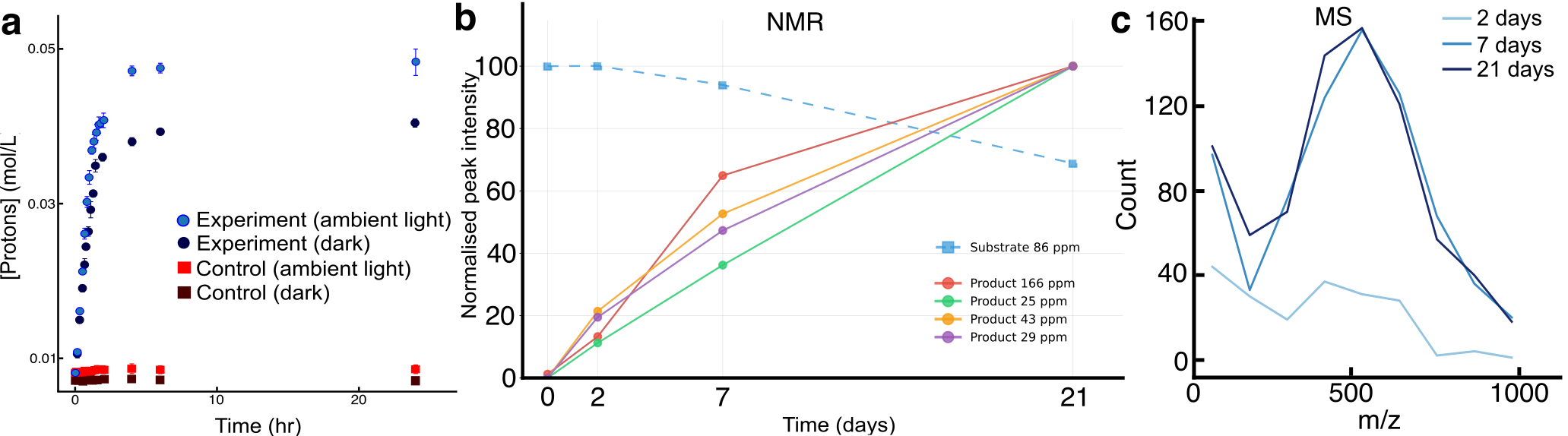}
\caption{\textbf{a}. For experiments performed both in ambient light and in the dark, compartment growth is accompanied by the production of protons in the reaction mixture. The controls do not show any significant change in pH over time.
\textbf{b}. Quantitative analysis of 1-D \textsuperscript{13}C Nuclear Magnetic Resonance (NMR) carbon shift peaks showing the time-course of substrate consumption and product formation, under controlled solar simulation conditions.
\textbf{c}. Mass spectrometry $m/z$ distribution plots for the time series experiments showing the mass count increases over time, under controlled solar simulation conditions.
}
\label{fig:s4}
\end{figure*}

\begin{figure*}
\centering
\includegraphics[width=0.7\textwidth]{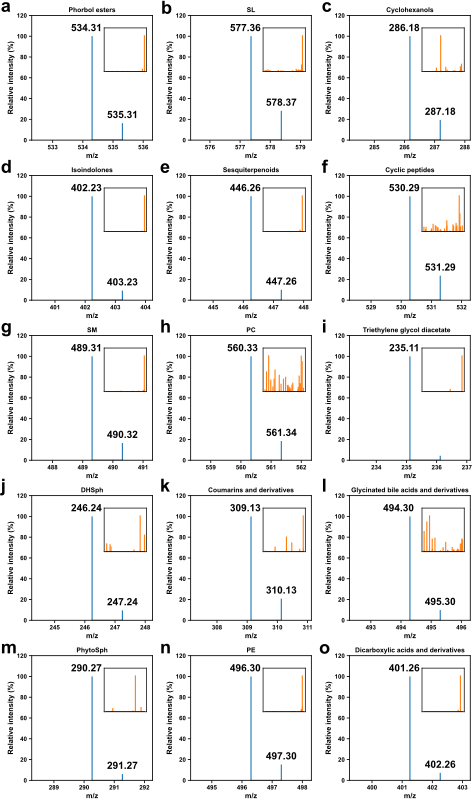}
\caption{Solar simulator laboratory experiment: Molecular identification for the protocell phase of the reaction mixtures spiked with 1 \% \textsuperscript{13}C-formaldehyde showcasing the major M (\textsuperscript{12}C peak) and its corresponding M+1 (\textsuperscript{13}C peak) for the following molecular species/types: \textbf{a}. Phorbol esters \textbf{b}. SL \textbf{c}. Cyclohexanols \textbf{d}. Isoindolones \textbf{e}. Sesquiterpenoids \textbf{f}. Cyclic peptides \textbf{g}. SM \textbf{h} PC \textbf{i}. Triethylene glycol diacetate \textbf{j}. DHSph \textbf{k}. Coumarin derivatives \textbf{l}. Glycinated bile acid derivatives \textbf{m}. PhytoSph \textbf{n}. PE \textbf{o} Dicarboxylic acid derivatives. Insets show the MS2 data for each of the species. All details of the MS analysis are summarised in Table~\ref{tab_lab}.
}
\label{fig:s4_lab}
\end{figure*}

\begin{figure*}
\centering
\includegraphics[width=0.8\textwidth]{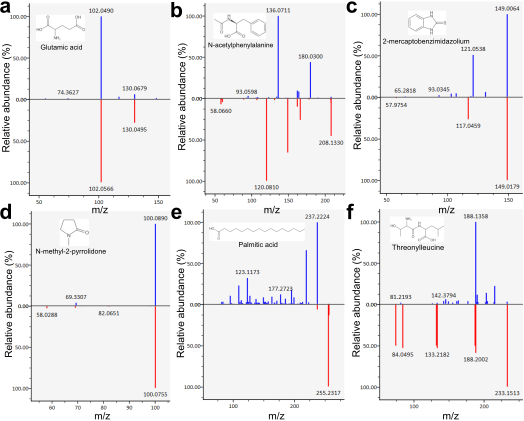}
\caption{Natural day-night cycling experiment: Molecular identification for the protocell phase of the reaction mixtures incubated under natural day-night cycles, showing various molecular species: \textbf{a}. Glutamic acid \textbf{b}. N-acetylphenylalanine \textbf{c}. 2-mercaptobenzimidazolium \textbf{d}. Palmitic acid \textbf{e}. Threonylleucine. For each of the species, both the MS1 and MS2 information is shown (in blue, top) and compared to reference database (in red, bottom). All details of the MS analysis are summarised in Table~\ref{tab_daynight}.
}
\label{fig:s4_sunlight}
\end{figure*}

\begin{figure*}
\centering
\includegraphics[width=0.8\textwidth]{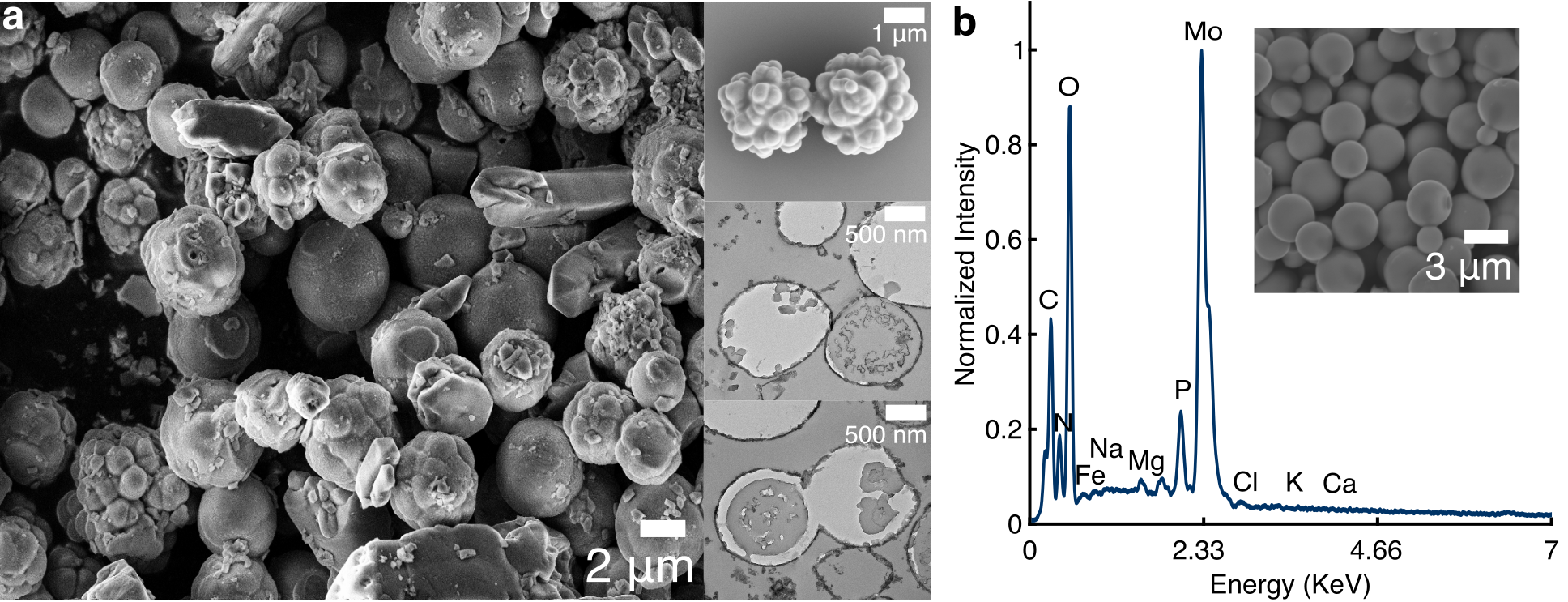}
\caption{\textbf{a} SEM images (left and top right panel) and TEM images (middle right and bottom right panel) of the laboratory protocells, particularly revealing their heterogeneous nature.
\textbf{b} EDX measurements of the protocells formed in reaction mixtures including additional components (sodium, magnesium, potassium and calcium salts), 24 hours post reaction initiation. Inset shows SEM image for the same.
}
\label{fig:s5}
\end{figure*}

\end{document}